\documentclass[12pt]{article}

\usepackage{amssymb,amsfonts,amsmath,mathtools,amsthm}
\usepackage[nohead]{geometry}
\usepackage[singlespacing]{setspace}
\usepackage[bottom]{footmisc}
\usepackage{indentfirst}
\usepackage{endnotes}
\usepackage{graphicx}%
\usepackage{rotating}
\usepackage{xcolor}
\usepackage{enumerate}
\RequirePackage{etex}
\usepackage[authoryear]{natbib}
\usepackage{hyperref}
\usepackage{undertilde}
\usepackage{tikz-network}
\usepackage{tikz}
\usepackage{caption,lipsum}
\usepackage{subcaption}
\usepackage{appendix}
\usepackage{pgfplots}
\pgfplotsset{compat=1.16}

\usetikzlibrary{arrows,shapes,backgrounds,automata, patterns, patterns.meta, decorations.pathreplacing}
\hypersetup{
	colorlinks=true,
	linkcolor=blue,
	citecolor=blue,
	filecolor=magenta,
	urlcolor=cyan,
}
\newtheorem{theorem}{Theorem}

\newtheorem{corollary}{Corollary}
\newtheorem{definition}{Definition}
\newtheorem{assumption}{Assumption}

\theoremstyle{definition}

\newtheorem{lemma}[theorem]{Lemma}
\newtheorem{proposition}{Proposition}

\newcommand\Tau{\mathcal{T}}

\renewcommand\thesubsection{\Alph{subsection}}

\newcommand{\added}[1]{#1}

\begin{document}
\title{A Model of Polarization on Social Media
\thanks{We would like to thank Sebastian Bervoets, Francis Bloch, Ugo Bolletta, Yann Bramoull\'{e}, Micael Castanheira, Fr\'{e}d\'{e}ric Dero\"{i}an, Matt Elliott, Garance Genicot, Sanjeev Goyal, Georg Kirchsteiger, Ana Mauleon, Dotan Persitz, Agnieszka Rusinowska, Leeat Yariv, Junjie Zhou and the participants of the Young Academics Network Conference at Cambridge-INET, SING16, $2^{nd}$ AMSE Summer School, the $7^{th}$ Annual Conference on Network Science and Economics, BINOMA 2022, CTN 2022, SAET 2024 and LAGV 2024 for their useful comments and suggestions. Merlino gratefully acknowledges the financial support of the FWO (grant G029621N), FNRS (grant T.0289.26), and the National Bank of Belgium. The usual disclaimers apply.}}
\author{Patrick Allmis\thanks{Faculty of Economics \& Janeway Institute, University of Cambridge, Sidgwick Ave, Cambridge CB3 9DD, United Kingdom; E-mail: pa509@cam.ac.uk.}
\\ \normalsize \textit{University of Cambridge}
\\ \\
Luca Paolo Merlino\thanks{ECARES, Universit\'{e} libre de Bruxelles, Ave. F.D. Roosevelt 50, CP 114/04, Belgium; email: luca.paolo.merlino@ulb.be.}
\\ \normalsize \textit{ECARES, Universit\'{e} libre de Bruxelles}
}

\date{February 15, 2026}

\maketitle

\sloppy

\onehalfspacing
\thispagestyle{empty}
\setcounter{page}{0}

\vspace{-.8cm}
\begin{abstract}
\noindent We develop a model of social media in which users produce different types of content and choose whom to follow. Even when abstracting from algorithmic bias, linking costs shape networks and polarization. In the welfare-maximizing equilibrium, lower linking costs can raise welfare but also increase exposure to extreme content, while very low costs reduce welfare and heighten polarization by discouraging moderate contributors. Policies that incentivize content provision can generate large welfare gains by changing who produces information, whereas link subsidies or attention reallocation mainly affect exposure and have limited welfare impact. These insights help explain why exposure-based interventions on social media platforms often yield ambiguous effects on polarization.
\end{abstract}


\smallskip


\onehalfspacing
\pagebreak

Social media have become a central gateway to political information, shaping how citizens form opinions and engage in public debate. By vastly expanding the set of potential information providers, digital platforms have reshaped how political content is produced, discovered, and consumed. In both the US and the UK, social media and video-sharing platforms now rival television as sources of news, and they are the primary news channels for the majority of young adults \citep{ofcom2024news,newman2025digital}. At the same time, they are widely viewed as key contributors to rising political polarization and declining social cohesion \citep{sunstein2018republic}.

Indeed, algorithms and selective following can limit exposure to cross-cutting views on platforms like Facebook \citep{bakshy2015exposure,braghieri2024article}, and feed design changes can shift exposure to ideologically slanted sources \citep{levy2021social}. Partisan content can influence attitudes \citep{munger2022political,kim2015does}, and its reach and effects increase with mobile internet access \citep{melnikov2025mobile}. Yet, many Twitter/X users still encounter diverse perspectives \citep{barbera2015birds,eady2019how}, and experiments modifying feeds often found negligible or negative effects on polarization \citep{bail2018,levy2021social,allcott2024effects}. These mixed patterns suggest that polarization need not rely on biased algorithms or platform manipulation, but can instead emerge endogenously from users' choices about whom to follow.

These findings highlight a key gap in understanding online polarization: when users can access abundant sources, a central driver of polarization is whom people choose to follow. These choices—shaped by preferences and the opportunity cost of acquiring content directly versus via other users—determine which information reaches them. Understanding when social media amplifies or dampens polarization (and the implications for welfare) therefore requires modeling both information demand and the networks shaping information production and flow.

To this end, we propose a model in which information is a local public good: each player can produce content that benefits anyone who follows them. Players make two types of decisions—how much content to produce and whom to follow. Therefore, a player's consumption depends on her own production, as well as the content produced by those she follows. Producing content is costly, reflecting the effort required to create posts, while following others is also costly, capturing the time, attention, and effort needed to discover and consume content.

Players differ in their relative preferences for two types of content—for example, left- versus right-leaning news. A player's type, ranging from 0 to 1, captures how strongly she prefers one type of content over the other: players with extreme preferences (types near 0 or 1) demand mostly one type, while moderates (types near 1/2) value both more evenly. Types also differ in the total amount of content they want to consume of both goods. This total demand may be higher for players with extreme preferences (types near 0 or 1) or for moderates (types near 1/2). We summarize this difference with a single measure, which we call ``zeal'': positive zeal means extremists demand more total content than moderates, while negative zeal means moderates demand more. Since following another player provides access to all the content that player produces, the benefit of linking to an extremist versus a moderate depends directly on zeal. Crucially, we abstract from algorithmic bias and amplification: polarization arises solely from players' production and following choices.

For concreteness, consider political information along a left–right spectrum. Users either gather information directly or rely on content produced by those they follow. Partisans value ideologically aligned content more, while moderates value both sides more evenly. Whether extremists or moderates consume more information in isolation—captured by zeal—shapes which types may become sources others follow.\footnote{Our insights extend to settings in which similar forces operate, e.g., knowledge-sharing networks \citep{shafiq2013identifying} or policy experimentation across jurisdictions \citep{callander2015experimentation}.}

In the model, players tend to connect to others with similar tastes because they demand similar information bundles, but also to those producing a lot. These two forces shape the network of information spillovers. This network, together with the induced pattern of content production and consumption, determines polarization, measured as the dispersion in information consumed across players \citep{esteban1994measurement}: greater divergence in consumption implies more polarization.

Nash equilibria take two forms: \textit{independent} equilibria, where some players provide their preferred information and others link to them in a bipartite structure, and \textit{collaborative} equilibria, where two contributors (at least one moderate) are linked and jointly supply information in a core–periphery pattern. Independent equilibria may feature extreme contributors and high polarization, while collaborative ones sustain moderate contributors and integrated networks

An independent equilibrium always exists, while a collaborative one requires sufficiently low zeal—ensuring moderate players provide enough total information—and intermediate linking costs. Even if multiple equilibria may arise, the fraction of agents who provide information to others remains bounded as the population grows—the so-called ``law of the few'' \citep{galeotti2010law}. Consistently, follower and activity distributions on social media such as Facebook, X and TikTok display skewed patterns \citep{shafiq2013identifying,twitter,tiktok2024us}.

To derive novel comparative statics, we focus on the equilibrium with the highest welfare. When zeal takes intermediate values, the welfare-maximizing equilibrium is collaborative. The intuition is as follows: if zeal is too low, moderate players already provide abundant public goods, so welfare is maximized when one moderate player serves as the sole network contributor, which is an independent equilibrium. Conversely, if zeal is too high, extremists provide so much more total information that a moderate contributor cannot supply enough to be valuable source.

Our model highlights that reducing linking costs—through easier following, recommendations, or interface design—can have non-monotonic effects on welfare and polarization. When links are expensive, extreme-taste players dominate as information providers. As costs fall, more users connect to them, increasing both welfare and polarization. At intermediate costs, moderates emerge as contributors, improving welfare while reducing polarization. But when costs fall further, these moderates stop producing and instead link to extreme sources, reducing welfare and deepening polarization. Polarization can thus emerge from design choices lowering linking costs—such as frictionless following or recommendation systems—even without algorithmic bias.

We derive how platforms should target incentives to boost public-good provision. With small budgets, supporting high-output contributors who already generate spillovers is optimal; with larger budgets, incentives should favor creators whose content attracts diverse audiences, even if they are less prolific. These predictions align with observed social media incentive programs, such as TikTok's Creator Rewards Program \citep{tiktok2025} and YouTube’s increasing emphasis on engagement and originality \citep{youtube}. By contrast, policies that subsidize links or reallocate attention—through recommendation algorithms or design choices that ease following—primarily reshuffle exposure rather than change who produces content, and thus typically deliver limited welfare gains. This helps explain why exposure-based interventions often yield ambiguous effects on polarization.

Taken together, our analysis demonstrates that polarization can emerge without any algorithmic bias: even when platforms treat users symmetrically, endogenous production and following choices generate polarized outcomes. This challenges the view that polarization stems primarily from algorithmic manipulation and shows why interventions targeting exposure alone cannot address the underlying mechanism.

A growing literature studies information and polarization on social media. \cite{acemoglu2024model} show how platform incentives amplify false content, while \cite{della2025affective} study the role of affective polarization and media bias. These papers focus on social learning conditional on exposure. We take a complementary approach, abstracting from learning to study how following and content production choices—shaped by linking incentives—drive polarization.

We model social media content creation as a public good, contributing to research on how group heterogeneity shapes provision and polarization. Prior work shows that homogeneity encourages contributions \citep{alesina1999public}; this can magnify small differences in endowments and drive sorting into similar communities \citep{benabou1996equity,durlauf1996theory}. Recent causal evidence finds more nuanced effects: integration and public good provision rise in highly fractionalized communities but fall in polarized ones \citep{bazzi2019unity}. We contribute by showing how taste heterogeneity affects polarization through its impact on content creation and following decisions.

Relatedly, \citet{baccara2013homophily,baccara2016choosing} study how players with heterogeneous preferences form groups in which all members contribute. We instead analyze the non‑cooperative, one‑sided formation of information networks, where a small set of users produces most of the content consumed by others, and where linking patterns—rather than group membership—drive polarization.

Finally, our paper builds on the literature on local public goods in networks. With a single public good on an exogenous network, Nash equilibria are specialized: some agents provide the public good, while others rely on their provision \citep{bramoulle2007public}. This insight extends to multiple goods when the network is fixed \citep{Walsh2019multi-layer}. We show that allowing the network to form endogenously fundamentally changes the equilibrium structure. As in \citet{galeotti2010law}, links are formed non‑cooperatively, but by introducing heterogeneous preferences \citep{kinateder2017public} and multiple goods, we uncover the trade‑off between linking to similar users and to prolific ones.\footnote{Networks can also be modeled non-cooperatively via socialization effort \citep{cabrales2011social,galeotti2014endogenous,merlino2014} or random opportunities \citep{konig2014nestedness}. \cite{sadler2024games} combine Nash equilibrium in actions and pairwise stability in links. Also, in our model spillovers flow in one network, complementing multigraph approaches \citep[e.g.,][]{joshi2020multigraph}.} For social media platforms, this implies that interventions targeting algorithms that reshape who follows whom—without altering users' incentives to create content—are unlikely to generate large welfare gains.

The paper proceeds as follows. Section \ref{model} introduces the model. Section \ref{equilibrium} characterizes the equilibria. Section \ref{largesocieties} studies welfare, polarization, and policies. Section \ref{discussion} shows that our insights are robust to benefits from receiving links, cost heterogeneity, and other extensions. Section \ref{conclusion} concludes. All proofs are in the appendix.

\section{Model}\label{model}

Let $N=\{1,2,...,n\}$ denote the set of players, where $i$ denotes the typical player and $n\geq 3$. Players benefit from their own contributions and those of the players they link to.\footnote{\added{We assume contributions of a given information good are perfect substitutes across sources. Following the literature on local public goods \citep[e.g.,][]{bramoulle2007public}, this focuses on total content rather than its source. Results are qualitatively robust to imperfect substitutability.}} In particular, player $i$ chooses her contributions to two goods, $x_{i}\in X=[0,\;+\infty)$ and $y_{i}\in Y=[0,\;+\infty)$. We say player $i$ is active in $x$ ($y$) if $x_{i}>0$ ($y_{i}>0$). \added{In our main application, the two public goods represent two types of information—for instance, left- and right-leaning content. Alternatively, one can view the setup as involving a single information good with two fixed, exogenous characteristics—such as ideological slant and depth of coverage. Evidence from social media shows that article-level slant and content characteristics shape consumption patterns and polarization \citep{braghieri2024article}, supporting this abstraction. Focusing on a fixed, two-dimensional set of goods provides the simplest environment to study how heterogeneity in tastes shapes polarization, while abstracting from endogenous content creation across categories—a reasonable approximation in settings where content domains or topics are well established.}\footnote{\added{More generally, the information good could have more than two characteristics, or there could be multiple distinct goods. In both cases, equilibrium structures remain qualitatively similar.}}

Player $i$ also chooses which links to establish: we denote $g_{ij}=1$ if player $i$ links to player $j$ for all $j\in N$. In doing so, player $i$ accesses any public good provided by $j$, i.e., $(x_j,y_j)$. As in \cite{bala2000noncooperative}, if $g_{ij}=1$, player $i$ links to $j$ at a cost $k>0$. We denote the resulting directed network ${\bf g}$ by an adjacency matrix in which each line represents a row vector, ${\bf g_{i}}=(g_{i1},...,g_{ii-1},g_{ii+1},...,g_{in})$, where $g_{ij}\in \{0,1\}\text{ for all } j\in N\setminus\{i\}$ and ${\bf g_{i}} \in G_{i}=\{0,1\}^{n-1}$. In our model link formation is one-sided and non-cooperative, and spillovers flow one-way: from the player receiving the link to the one establishing it. \added{These assumptions capture social media like TikTok, X/Twitter or YouTube, where following a user allows one to see the content they post. Consuming others' content is costly for those establishing the link, but not for those receiving it. The cost of linking captures, in reduced form, uniform search frictions: the time and effort required to find and connect to a user.}

Let $N_{i}({\bf g})=\{j\in N:g_{ij}=1\}$ be the set of players $i$ establishes a link to. We denote by $\eta_{i}({\bf g})=\vert N_{i}({\bf g}) \vert$ the number of players $i$ links to. Let $I({\bf g})=\{i\in N: g_{ij}=g_{ji}=0\;\text{ for all }\;j\in N\}$ denote the set of \textit{isolated players}. Denote the set of \textit{periphery players} who only sponsor links by $P({\bf g})=\{i\in N: g_{ij}=1 \text{ for some } j\in N \text{ and } g_{ji}=0 \text{ for all } j\in N\}$. Finally, let $C({\bf g})=\{i\in N: g_{ji}=1 \text{ for some } j\in N\}$ be the set of players who receive at least one link in ${\bf g}$; we refer to such players as the \textit{network contributors} because their contributions generate spillovers in their local network.

\begin{figure}[ht]
    \centering
    \begin{subfigure}[b]{0.4\linewidth}
    \begin{tikzpicture}[
	contributor/.style={diamond,draw,thick, fill=blue!20, scale=0.7},
	periphery/.style={circle,draw,thick, scale=0.7, fill=yellow!20},
	core/.style={regular polygon,regular polygon sides=5,draw,thick, fill=green!20, scale=0.7},
	]
	\draw[->,thin] (0,0)--(5,0)node[right]{$t$};
    \node [below] ($0$) at (0,-.1) {0};
    \node [below] ($1$) at (5,-.1) {1};
	\path[draw,thin] (0,-0.2) edge node {} (0,0.2);
	\node[contributor] at (5,3) (1){1};
	\node[contributor] at (0,3) (0){0};
	\node[contributor] at (4,3) (i){i};
	\node[contributor] at (1,3) (j){j};
	\node[contributor] at (3.14,3) (h){h};
	\node[contributor] at (1.86,3) (l){l};
	\node[periphery] at (0.5,0.3) (0'){};
	\node[periphery] at (4.5,0.3) (1'){};
	\node[periphery] at (2.64,0.3) (h'){};
	\node[periphery] at (2.36,0.3) (l'){};
	\node[periphery] at (3.5,0.3) (i'){};
	\node[periphery] at (1.5,0.3) (j'){};
	\node[periphery] at (0.25,0.15) (0''){};
	\node[periphery] at (4.75,0.15) (1''){};
	\node[periphery] at (2.89,0.15) (h''){};
	\node[periphery] at (2.11,0.15) (l''){};
	\node[periphery] at (3.75,0.15) (i''){};
	\node[periphery] at (1.25,0.15) (j''){};
	\draw[->,thin] (1')--(1);
	\draw[->,thin] (0')--(0);
    \draw[->,thin] (i')--(i);
	\draw[->,thin] (j')--(j);
    \draw[->,thin] (h')--(h);
	\draw[->,thin] (l')--(l);
	\draw[thin] (1'')--(1);
	\draw[thin] (0'')--(0);
    \draw[thin] (i'')--(i);
	\draw[thin] (j'')--(j);
    \draw[thin] (h'')--(h);
	\draw[thin] (l'')--(l);
	\end{tikzpicture}
	\caption{A bipartite graph}\label{bipartite}
	\end{subfigure}
	\begin{subfigure}[b]{0.4\linewidth}
    \begin{tikzpicture}[
	contributor/.style={diamond,draw,thick, fill=blue!20, scale=0.7},
	periphery/.style={circle,draw,thick, scale=0.7, fill=yellow!20},
	core/.style={regular polygon,regular polygon sides=5,draw,thick, fill=green!20, scale=0.7},
	]
	\draw[->,thin] (0,0)--(5,0)node[right]{$t$};
	\path[draw,thin] (0,-0.2) edge node {} (0,0.2);
    \node [below] ($0$) at (0,-.1) {0};
    \node [below] ($1$) at (5,-.1) {1};
	\node[periphery] at (5,0.6) (1){1};
	\node[periphery] at (0,0.9) (0){0};
	\node[core] at (1.5,2.1) (j){j};
	\node[periphery] at (0.74673,0.452) (q'){};
	\node[periphery] at (2.5,0) (m){};
	\node[core] at (4,2.4) (i){i};
	\node[periphery] at (4.18687,0.112) (q){};
	\draw[->,thin] (j)--(i);
	\draw[->,thin] (m)--(j);
    \draw[->,thin] (m)--(i);
	\draw[->,thin] (i)--(j);
    \draw[->,thin] (q)--(i);
	\draw[->,thin] (q)--(j);
    \draw[->,thin] (q')--(i);
	\draw[->,thin] (q')--(j);
	\draw[->,thin] (1)--(i);
	\draw[->,thin] (0)--(j);
	\end{tikzpicture}
	\caption{A core-periphery graph}\label{coreperiphery}
    \end{subfigure}
\caption{Two examples of networks.}\label{fig:networks}
\begin{minipage}{\linewidth}
\footnotesize
\emph{Notes:} Players are represented by nodes (ordered by their type from $0$ to $1$ on the horizontal axis; circles represent players who do not receive links) and links by arrows. The vertical axis represents players' total contribution to the two public goods.
\end{minipage}
\end{figure}
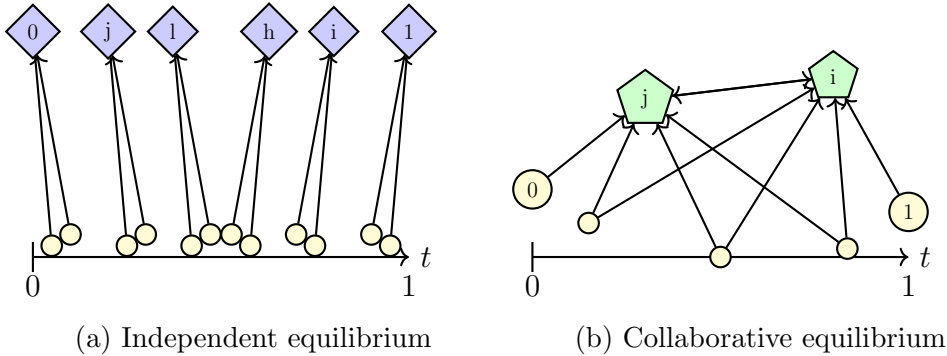

\added{Departing from the usual definitions, we define the following structures on the directed network ${\bf g}$.} A graph is \textit{bipartite} if all players \added{who have at least one link} can be partitioned into two sets where no players within the same set are linked, but every player in one set is linked to at least one player in the other set; \added{isolated players are permitted}. Formally, for any players $i,j\in P({\bf g})$ or $i,j\in C({\bf g})$, $g_{ij}=g_{ji}=0$, while for any $i\in P({\bf g})$, there exist $l\in C({\bf g})$ such that $g_{il}=1$. Figure \ref{bipartite} depicts an example of a bipartite graph. A \textit{core-periphery graph} consists of two groups of players, the \textit{core} and the \textit{periphery}, such that core players are linked among themselves, while periphery players link to at least one core player. Formally, for any players $i,j\in P({\bf g})$, $g_{ij}=g_{ji}=0$, while for any players $l,m\in C({\bf g})$, $\max\{g_{lm},g_{ml}\}=1$. Moreover, for any $i\in P({\bf g})$, there exists $l\in C({\bf g})$ such that $g_{il}=1$. Figure \ref{coreperiphery} depicts an example of a core-periphery graph. The network $\mathbf{g}$ is a \textit{star} if $\vert C(\mathbf{g})\vert=1$ and $\vert P(\mathbf{g})\vert=n-1$.\footnote{Note that the star is the only network that is both a bipartite graph and a core-periphery graph.}

Let $S_{i}=X\times Y\times G_{i}$ denote the strategy space of player $i$ and $S=S_{1}\times ...\times S_{n}$ the set of strategies of all players. A strategy profile ${\bf s}=({\bf x}, {\bf y}, {\bf g})\in S$ 
defines each player's contributions and the links they create. Under the strategy profile ${\bf s}=({\bf x},{\bf y},{\bf g})$, let $\bar{x}_{i}({\bf g})=\sum_{j\in N_{i}({\bf g})} x_{j}$ and $\bar{y}_{i}({\bf g})= \sum_{j\in N_{i}({\bf g})} y_{j}$ denote the public good $x$ and $y$, respectively, received by player $i$ via their neighbors under ${\bf s}$.

Player $i$'s utility is given by
\begin{equation}\label{payoffs}
U_{i}({\bf s})=t_{i}f(x_{i}+\bar{x}_{i}({\bf g}))+(1-t_{i})f(y_{i}+\bar{y}_{i}({\bf g}))-(x_{i}+y_{i}) c-\eta_{i}({\bf g})k,
\end{equation}
where $f(\cdot)$ denotes benefits from each good, $t_i \in [0,1]$ is $i$’s taste parameter (higher $t_i$ indicates stronger preference for good $x$), $c>0$ is the linear cost of contributing, and $k>0$ is the cost per link initiated.\footnote{In Section~\ref{discussion}, we relax cost homogeneity, allow weighted links and benefits from incoming links, and explore alternative spillover specifications to study intrinsic motivation and decay.}

Let $f(\cdot)$ be twice continuously differentiable, strictly concave, and increasing in $x$ and $y$. We assume that $\lim_{z\to 0} f'(z)>c$ and $\lim_{z\to\infty} f'(z)=m<c$ for all $i\in N$. \added{These assumptions ensure that, for any network ${\bf g}$, player $i$ has a well-defined optimal contribution bundle when isolated, with a unique interior solution for $t_i\in(0,1)$ and a corner solution satisfying $\hat{y}_i=0$ if $t_i=1$ and $\hat{x}_i=0$ if $t_i=0$.}

\begin{definition}\label{def:isolation}
\added{A player $i$'s \emph{isolation demand} is the bundle}
\[
(\hat{x}_i,\hat{y}_i)
:= \arg\max_{(x,y)\in X\times Y} \left\{ t_i f(x) + (1-t_i) f(y) - (x+y) c \right\},
\]
\added{which maximizes $i$'s utility when she receives no benefits from others' contributions.}
\end{definition}
\added{Isolation demand plays an important role in the analysis, as players’ incentives to form links depend on how these preferred bundles compare across individuals.}

\added{An implication of content being a local public good is that players who link to others benefit from their contributions without affecting the contributors’ incentives. As a result, a player may optimally reduce her own content provision and rely instead on content produced by users she follows. We refer to this behavior as \emph{free riding}.}

\added{Two comments are in order. First, each unit of $x_i$ or $y_i$ represents a piece of information. Consuming more units increases utility, and the additive terms $\bar{x}_i({\bf g})$ and $\bar{y}_i({\bf g})$ capture the content received by player $i$ through the users she follows on the platform. These variables therefore represent total exposure to each type of content, aggregating own production and content accessed via the network. While not relying on formal Bayesian learning, one can view this exposure as a reduced-form measure of the information environment faced by a user, which shapes polarization. This interpretation aligns with evidence that social media consumption often has entertainment and preference-driven components \citep{dellavigna2015media}.}

\added{Second, heterogeneous tastes over $(x,y)$ imply that users benefit most from following sources with similar preferences, as they provide an information mix close to their ideal content---i.e., their demand in isolation. Yet, users may follow highly active or specialized content producers even when preferences differ, trading off ideological similarity for access to larger amounts of a particular type of information.}

Each player $i\in N$ draws $t_{i}$ from the continuous cumulative distribution function $\Tau\sim [0,1]$. To simplify the exposition, we assume $t_{i}>t_{j}$ if $i>j$ for all $i,j\in N$, \added{and normalize so that the lowest and highest types satisfy $t_0=0$ and $t_1=1$}.\footnote{\added{Assuming $t_i \neq t_j$ for all $i,j$ is for expositional convenience; relaxing this assumption may yield additional equilibria due to indifference, which can be removed by considering strict Nash equilibria \citep{galeotti2010law,kinateder2017public}}.}

A strategy profile ${\bf s}^{\ast}=({\bf s}_{i}^{\ast}, {\bf s}_{-i}^{\ast})$ is a \textit{Nash equilibrium} if $U_{i}({\bf s}_{i}^{\ast},{\bf s}_{-i}^{\ast})\geq U_{i}({\bf s}_{i}',{\bf s}_{-i}^{\ast})$ for all ${\bf s}_{i}'\in S_i$ and $i\in N$. \added{Since linking determines whether users rely on their own production or on others, it is useful to distinguish two classes of equilibria.}
\begin{definition}\label{def:ind_collab}
\added{An equilibrium is \textbf{independent} if contributors form no links with one another, i.e.\ $g^\ast_{ij}=0$ for all $i,j\in C(\mathbf{g}^\ast)$, so each contributor supplies all the content they consume. It is \textbf{collaborative} otherwise, meaning that some contributors link to others to obtain part of their content.}
\end{definition}

Welfare is the sum of the utilities of all players $\sum_{i\in N}U_{i}({\bf s})$. The \textit{welfare-maximizing equilibrium} $\mathbf{s}^W$ is the Nash equilibrium such that $\sum_{i\in N}U_i(\mathbf{s}^W)\ge \sum_{i\in N}U_i(\mathbf{s}^\ast)$ for all $\mathbf{s}^\ast$.

For a strategy profile $\mathbf{s}$, let polarization be $\rho({\bf s})=\sum_{i\in N}\sum_{j\in N}(\vert (x_i+\bar{x}_i({\bf g}))- (x_j+\bar{x}_j({\bf g}))\vert+\vert (y_i+\bar{y}_i({\bf g}))- (y_j+\bar{y}_j({\bf g}))\vert)$ \citep{esteban1994measurement}. This notion captures how dissimilar players' consumption bundles are under a given strategy profile. \added{In our context, it captures the extent to which users consume different amounts of left- or right-leaning content (or content with different characteristics such as slant or depth), reflecting divergence in information consumption across individuals.}

\section{Equilibrium characterization}\label{equilibrium}

To characterize the equilibria of the game, we first extend a result of \cite{bramoulle2007public} on equilibrium consumption to players with different valuations for two goods. In particular, utility maximization implies that they consume at least as much as in isolation, and exactly so for any public good they provide. Players who achieve a lower level of public goods than they would in isolation have an incentive to contribute until they reach that level. They will consume more only if they completely free ride on the contributions of others. Formally:
\begin{lemma}\label{lemma:BK}
In any equilibrium ${\bf s}^{\ast}=({\bf x}^{\ast},{\bf y}^{\ast},\mathbf{g}^\ast)$,
\begin{enumerate}[{(}i{)}]
\item $x^{\ast}_{i}+\bar{x}^{\ast}_{i}(\mathbf{g}^\ast)\geq \hat{x}_{i}$ and $y^{\ast}_{i}+\bar{y}^{\ast}_{i}(\mathbf{g}^\ast)\geq \hat{y}_{i}$, for all $i\in N$
\item if $x^{\ast}_{i}>0$, then $x^{\ast}_{i}+\bar{x}^{\ast}_{i}(\mathbf{g}^\ast)=\hat{x}_{i}$ and if $y^{\ast}_{i}>0$, then $y^{\ast}_{i}+\bar{y}^{\ast}_{i}(\mathbf{g}^\ast)=\hat{y}_{i}$.
\end{enumerate}
\end{lemma}

In our framework, heterogeneity in the valuations for the public goods generates different demands for them. The gap in total public goods demand between players with extreme and moderate tastes plays a key role in our analysis. We measure this difference by \textit{zeal}, $z(f,c)$:
\begin{eqnarray}\label{zeal}
    z(f,c)=\frac{(\hat{x}_0+\hat{y}_0)-(\hat{x}_{\frac{1}{2}}+\hat{y}_{\frac{1}{2}})}{\max\{(\hat{x}_0+\hat{y}_0),(\hat{x}_{\frac{1}{2}}+\hat{y}_{\frac{1}{2}})\}}.
\end{eqnarray}
\added{Zeal captures how total information demand differs between moderate and extreme players. Consider the taste parameter $t_i$ as an index of how extreme the taste of a player is: if $t_i$ is close to $0$ or $1$, player $i$ is an \textit{extremist} of good $y$ or $x$, demanding much more of one good than the other; if $t_i$ is close to $1/2$, $i$ is a \textit{moderate}, valuing both goods more evenly. When $z(f,c)>0$, extremists demand more \textit{total} information, whereas moderates do when $z(f,c)<0$. Formally, zeal is the normalized difference in total information demand between the most extreme type (player $0$) and the most moderate type (player $1/2$), so that $z(f,c)\in[-1,1]$.}\footnote{Zeal is exogenous---it depends on isolation demands. Comparing the demands of player $1/2$ and player $1$ yields the same measure of zeal, \added{since total isolation demand is symmetric around $1/2$.}}

For some results, \added{we assume that the curvature of benefits from consuming a given type of content is monotone across the type space.}
\begin{assumption}\label{ass:zeal}
Assume that $f(\cdot)$ is three times continuously differentiable and either $f'''(v)\geq 0$ or $f'''(v)\leq 0$ for all $v\in[0,+\infty)$.
\end{assumption}
\added{Specifically, the benefit function $f$, which applies to both goods, and its higher-order curvature governs how marginal benefits change with consumption of each good. Intuitively, $f'''$ determines how quickly marginal benefits decline as consumption of one good increases. If $f'''\geq 0$, marginal benefits fall more slowly at higher consumption levels, while if $f'''\leq 0$, marginal benefits fall more quickly.}

The following lemma describes how zeal shapes players' isolation demands.
\begin{lemma}\label{lemma:zeal}
A player's isolation demand for a public good satisfies:
\begin{enumerate}[{(}i{)}]
    \item \added{$\hat{x}_i$ (resp.\ $\hat{y}_i$) is strictly increasing (resp.\ strictly decreasing) in $t_i$.}
    \item \added{Under Assumption~\ref{ass:zeal}, total isolation demand of a player $i$, $\hat{x}_i+\hat{y}_i$, is strictly convex in $|t_i-\tfrac12|$ when $z(f,c)>0$, and strictly concave when $z(f,c)<0$.}
\end{enumerate}
\end{lemma}
\added{Part (i) is immediate: as a player’s taste for good $x$ increases, she allocates more consumption to $x$ and less to $y$. 
Part (ii) reflects how the curvature of the benefit function determines whether total isolation demand is higher for extreme or moderate players, a difference summarized by zeal. 
When marginal benefits decline slowly ($f'''> 0$), concentrating consumption on a single good entails a relatively small marginal utility loss. As a result, players with more extreme tastes demand substantially more of their preferred good than they reduce consumption of the other, so their total isolation demand increases with distance from $t_i=\tfrac12$, making it strictly convex in $|t_i-\tfrac12|$. The opposite happens when marginal benefits decline quickly ($f'''< 0$). When $f''' =0$, total isolation demand is constant and zeal equals zero.}

\added{We now characterize the structure of the Nash equilibria of the game.}
\begin{proposition}\label{prop:charact}
\added{A Nash equilibrium always exists. Moreover, every Nash equilibrium takes one of two forms:}
\begin{enumerate}[{(}i{)}]

    \item \added{\textit{Independent equilibrium.}
    The network $\mathbf{g}^\ast$ is a bipartite graph (possibly with isolated players) in which all network contributors sponsor no links. If there are at least three network contributors, all other players sponsor at most one link.}

    \item \added{\textit{Collaborative equilibrium.}
    The network $\mathbf{g}^\ast$ is a core–periphery graph with exactly two network contributors in the core, and all other players in the periphery linking to at least one of them.}

\end{enumerate}
\end{proposition}

\added{The proof of existence is constructive. Starting from the players with the largest isolation demands, one can iteratively add all profitable links: players link to whoever provides the most-valued content, while contributors never sponsor links themselves. This process always terminates, yielding a bipartite network in which contributors supply their entire preferred bundle and all other players either free ride on them or remain isolated. Hence, this equilibrium is independent, as described in part (i).}

\added{As an illustration of part (i), Figure \ref{fig-prop2ii} shows an independent equilibrium under a logarithmic benefit function (so zeal is zero and the sum of the isolation demand of the two goods is constant across types). The resulting network is bipartite: network contributors sponsor no links and supply their entire preferred bundles, while all other players either free ride on them or remain isolated. When there are at least three network contributors, the fact that every other player sponsors at most one link implies that players sort into disjoint follower groups, each group having only one link, but to a different contributor.}

\begin{figure}[ht]
    \centering
    \begin{subfigure}[b]{0.32\linewidth}
    \centering
    \begin{tikzpicture}[
	contributor/.style={diamond,draw,thick, fill=blue!20, scale=0.6},
	periphery/.style={circle,draw,thick, scale=0.45, fill=yellow!20},
	transparentperiphery/.style={circle,draw,thick, scale=0.6, fill=white, transparent},
	core/.style={regular polygon,regular polygon sides=5,draw,thick, fill=green!20, scale=0.6},
	isolated/.style={regular polygon,regular polygon sides=7,draw,thick, fill=red!20, scale=0.6},
	]
	\path[draw,thick,below] (0,0) edge node {} (4.2,0);
	\path[draw,thick,left] (0,0) edge node {} (0,3.5);
	\node[contributor] at (4.2,3) (1){1};
	\node[contributor] at (0,3) (0){0};
	\node[isolated] at (2.73,3) (i){.65};
	\node[isolated] at (1.47,3) (j){.35};
	\node[periphery] at (3.57,.45) (p){.85};
	\node[isolated] at (2.1,3) (m){.5};
    
	\node[periphery] at (3.78,.3) (h){.9};
	\node[periphery] at (.42,.3) (k){.1};
	\node[periphery] at (.63,.45) (q){.15};
    \draw[->,thin] (k)--(0);
	\draw[->,thin] (h)--(1);
    \draw[->,thin] (q)--(0);
	\draw[->,thin] (p)--(1);
	\end{tikzpicture}
	\caption{$k=0.985$}
	\end{subfigure}
    \begin{subfigure}[b]{0.32\linewidth}
    \centering
    \begin{tikzpicture}[
	contributor/.style={diamond,draw,thick, fill=blue!20, scale=0.6},
	periphery/.style={circle,draw,thick, scale=0.45, fill=yellow!20},
	transparentperiphery/.style={circle,draw,thick, scale=0.6, fill=white, transparent},
	core/.style={regular polygon,regular polygon sides=5,draw,thick, fill=green!20, scale=0.6},
	isolated/.style={regular polygon,regular polygon sides=7,draw,thick, fill=red!20, scale=0.6},
	]
	\path[draw,thick,below] (0,0) edge node {} (4.2,0);
	\path[draw,thick,left] (0,0) edge node {} (0,3.5);
	\node[contributor] at (4.2,3) (1){1};
	\node[contributor] at (0,3) (0){0};
	
	\node[periphery] at (2.73,0.45) (i){.65};
	\node[periphery] at (1.47,0.45) (j){.35};
	\node[periphery] at (.63,.45) (q){.15};
	\node[periphery] at (3.57,0.45) (p){.85};
	\node[contributor] at (2.1,3) (m){.5};
    
	\node[periphery] at (3.78,.3) (h){.9};
	\node[periphery] at (.42,.3) (k){.1};
    \draw[->,thin] (k)--(0);
	\draw[->,thin] (h)--(1);
	\draw[->,thin] (p)--(1);
	\draw[->,thin] (q)--(0);
    \draw[->,thin] (i)--(m);
	\draw[->,thin] (j)--(m);
	\end{tikzpicture}
	\caption{$k=0.95$}
	\end{subfigure}\begin{subfigure}[b]{0.32\linewidth}
    \centering
    \begin{tikzpicture}[
	contributor/.style={diamond,draw,thick, fill=blue!20, scale=0.6},
	periphery/.style={circle,draw,thick, scale=0.45, fill=yellow!20},
	transparentperiphery/.style={circle,draw,thick, scale=0.6, fill=white, transparent},
	core/.style={regular polygon,regular polygon sides=5,draw,thick, fill=green!20, scale=0.6},
	isolated/.style={regular polygon,regular polygon sides=7,draw,thick, fill=red!20, scale=0.6},
	]
	\path[draw,thick,below] (0,0) edge node {} (4.2,0);
	\path[draw,thick,left] (0,0) edge node {} (0,3.5);
	\node[contributor] at (4.2,3) (1){1};
	\node[contributor] at (0,3) (0){0};
    
	\node[periphery] at (3.78,.3) (h){.9};
	\node[periphery] at (.42,.3) (k){.1};
    \draw[->,thin] (k)--(0);
	\draw[->,thin] (h)--(1);
	\node[periphery] at (2.73,0) (i){.65};
	\node[periphery] at (1.47,0) (j){.35};
	\node[periphery] at (.63,0) (q){.15};
	\node[periphery] at (3.57,0) (p){.85};
	\node[periphery] at (2.1,0) (m){.5};
	\draw[->,thin] (i)--(1);
	\draw[->,thin] (i)--(0);
	\draw[->,thin] (j)--(1);
	\draw[->,thin] (j)--(0);
	\draw[->,thin] (p)--(1);
	\draw[->,thin] (p)--(0);
	\draw[->,thin] (q)--(1);
	\draw[->,thin] (q)--(0);
	\draw[->,thin] (m)--(1);
	\draw[->,thin] (m)--(0);
	\end{tikzpicture}
	\caption{$k=0.4$}
	\end{subfigure}\\
    \begin{tikzpicture}[
	contributor/.style={diamond,draw,thick, fill=blue!20, scale=0.6},
	periphery/.style={circle,draw,thick, scale=0.7, fill=yellow!20},
	core/.style={regular polygon,regular polygon sides=5,draw,thick, fill=green!20, scale=0.6},
	isolated/.style={regular polygon,regular polygon sides=7,draw,thick, fill=red!20, scale=0.6},
	]
	\centering
	\matrix [draw,below left] at (current bounding box.north west) {
  \node [contributor,label=right:Contributor] {}; &
  \node [periphery,label=right:Periphery] {}; &
  \node [isolated,label=right:Isolated] {};\\
};
    \end{tikzpicture}
\caption{Examples of independent equilibria.}\label{fig-prop2ii}
\begin{minipage}{\linewidth}
\smallskip
\footnotesize
\emph{Notes:} In this example, $f(\cdot)=\ln(\cdot)$ and $c=1$. The horizontal axis represents types ranging from $0$ to $1$, and the vertical axis represents total public good contributions of players. Nodes represent agents and indicate the corresponding type; arrows represent links.
\end{minipage}
\end{figure}

The \emph{collaborative equilibria} in part (ii) arise only when there are exactly two network contributors who maintain at least one link between them, forming a core–periphery structure. Each core player is the highest contributor to one of the two goods, and the link between them allows at least one of them to benefit from the other's provision. \added{No additional player can join the core: by Lemma \ref{lemma:BK}, any player who sponsors a link provides at most one public good, which precludes a third contributor from linking to either core player.} All remaining players free ride by linking to one or both core contributors.\footnote{\added{The equilibrium structure we identify differs from models with a single public good or with endogenous group formation. With one public good, equilibria are nested-split graphs, with contributors at the core and periphery players in nested neighborhoods \citep{kinateder2017public}. Here, collaborative equilibria feature exactly two contributors with non‑nested neighborhoods, while stars arise only when zeal is sufficiently negative (so moderates demand more total content than extremists). Unlike costly group‑formation models \citep{baccara2016choosing}, extremes share a group only when all players—including moderates—belong to the same group.}}

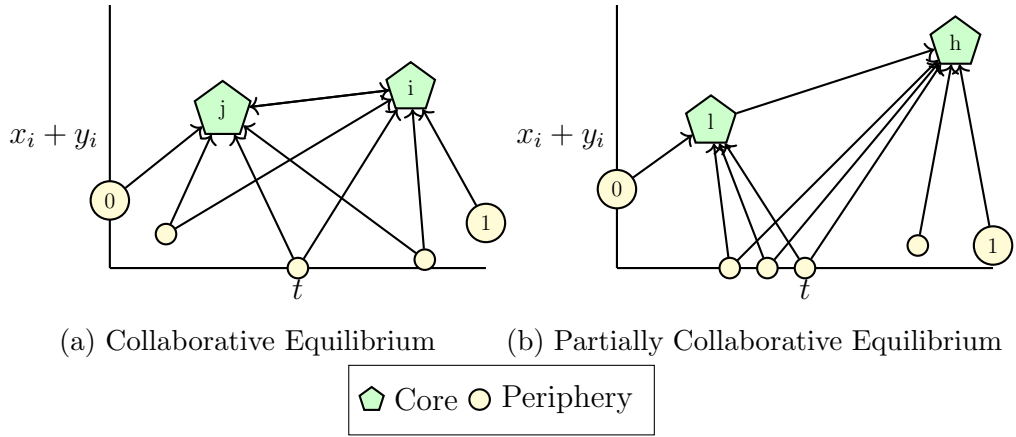
\begin{figure}[ht]
    \centering
	\begin{subfigure}[b]{0.4\linewidth}
    \begin{tikzpicture}[
	contributor/.style={diamond,draw,thick, fill=blue!20, scale=0.7},
	periphery/.style={circle,draw,thick, scale=0.45, fill=yellow!20},
	core/.style={regular polygon,regular polygon sides=5,draw,thick, fill=green!20, scale=0.7},
	]
	\path[draw,thick,below] (0,0) edge node {$t$} (5,0);
	\path[draw,thick,left] (0,0) edge node {$x_{i}+y_{i}$} (0,3.5);
	\node[periphery] at (5,1.05) (1){1};
	\node[periphery] at (0,1.05) (0){0};
	\node[periphery] at (.75,.6) (q){.15};
	\node[core] at (1.75,1.95) (j){.35};
	\node[periphery] at (2.5,0) (m){.5};
	\node[core] at (3.25,1.95) (i){.65};
	\node[periphery] at (4.25,0.6) (p){.85};
    \node[periphery] at (4.5,.75) (h){.9};
	\node[periphery] at (.45,.75) (k){.1};
	\draw[->,thin] (m)--(i);
    \draw[->,thin] (m)--(j);
	\draw[->,thin] (j)--(i);
	\draw[->,thin] (i)--(j);
	\draw[->,thin] (1)--(i);
	\draw[->,thin] (0)--(j);
	\draw[->,thin] (q)--(j);
	\draw[->,thin] (p)--(i);
	\draw[->,thin] (h)--(i);
    \draw[->,thin] (k)--(j);
	\end{tikzpicture}
 \caption{Collaborative Equilibrium}\label{fig:c}
    \end{subfigure}
	\begin{subfigure}[b]{0.4\linewidth}
    \begin{tikzpicture}[
	contributor/.style={diamond,draw,thick, fill=blue!20, scale=0.7},
	periphery/.style={circle,draw,thick, scale=0.45, fill=yellow!20},
	core/.style={regular polygon,regular polygon sides=5,draw,thick, fill=green!20, scale=0.7},
	]
	\path[draw,thick,below] (0,0) edge node {$t$} (5,0);
	\path[draw,thick,left] (0,0) edge node {$x_{i}+y_{i}$} (0,3.5);
	\node[periphery] at (5,.45) (1){1};
	\node[periphery] at (0,.9) (0){0};
	\node[core] at (.75,2.1) (q){.15};
	\node[periphery] at (1.75,0) (j){.35};
	\node[periphery] at (2.5,0) (m){.5};
	\node[periphery] at (3.25,0) (i){.65};
	\node[core] at (4.25,3) (p){.85};
    \node[periphery] at (4.5,0.15) (h){.9};
	\node[periphery] at (.5,.15) (k){.1};
	\draw[->,thin] (m)--(p);
    \draw[->,thin] (m)--(q);
	\draw[->,thin] (j)--(p);
	\draw[->,thin] (i)--(p);;
	\draw[->,thin] (j)--(q);
	\draw[->,thin] (i)--(q);
	\draw[->,thin] (1)--(p);
	\draw[->,thin] (0)--(q);
	\draw[->,thin] (q)--(p);
	\draw[->,thin] (h)--(p);
    \draw[->,thin] (k)--(q);
	\draw[->,thin] (k)--(p);
	\end{tikzpicture}
    \caption{Collaborative Equilibrium}\label{fig:pc}
    \end{subfigure}
    \begin{tikzpicture}[
	contributor/.style={diamond,draw,thick, fill=blue!20, scale=0.7},
	periphery/.style={circle,draw,thick, scale=0.7, fill=yellow!20},
	core/.style={regular polygon,regular polygon sides=5,draw,thick, fill=green!20, scale=0.7},
	]
	\matrix [draw,below left] at (current bounding box.north west) {
    \node [core,label=right:Core] {}; &
    \node [periphery,label=right:Periphery] {}; \\
};
    \end{tikzpicture}
\caption{Examples of collaborative equilibria.}\label{fig-prop2i}
\begin{minipage}{\linewidth}
\smallskip
\footnotesize
\emph{Notes:} In this example, $f(\cdot)=\ln(\cdot)$, $c=1$ and $k=0.4$. The horizontal axis represents types ranging from $0$ to $1$, and the vertical axis represents total public good contributions of players. Nodes represent agents and indicate the corresponding type; arrows represent links.
\end{minipage}
\end{figure}

Figure \ref{fig-prop2i} shows that not all types can be network contributors in a collaborative equilibrium. In the example, only types in the intermediate intervals $t_i\in[.18,.4)$ and $t_i\in(.6,.82]$ can form the two-player core—these types value each other’s contributions enough to link. More extreme types would not find it profitable to link, whereas more moderate types prefer to free ride on players close to $0$ or $1$. When only one of the two contributors free rides on the other, the core player who sponsors the link can instead be more extreme: for instance, in panel (b) of Figure \ref{fig-prop2i}, a player with type $t=0.8$ links to a more extreme contributor while still serving as a core player.

\added{The two equilibrium structures generate distinct patterns of polarization. Since the two core contributors must lie in the intervals where linking is profitable for at least one of them (Figure \ref{fig-prop2i}), at least one of them is necessarily moderate. Their content provision is therefore not highly polarized. Moreover, because all remaining players link to one or both core contributors, users' consumption bundles concentrate around these moderate sources, leading to substantial overlap in informational exposure and relatively low polarization.}

\added{In contrast, \emph{independent equilibria} can exhibit very high or very low polarization, depending on which types serve as network contributors and which users link to them. As shown in panel (a) of Figure \ref{fig-prop2ii}, if zeal is non-negative, high linking costs typically imply that only the most extreme types to contribute and that users link to at most one contributor. Left-leaning users follow the left extreme and right-leaning users follow the right extreme, yielding highly polarized consumption patterns. When linking costs are sufficiently cheap as in panel (c) of Figure \ref{fig-prop2ii}, many users link to both extremes, consumption bundles are similar, so polarization is low.}

\added{For intermediate linking costs, additional links become profitable, allowing some moderate types to free ride on contributors and potentially altering which types are network contributors (panel (b) of Figure \ref{fig-prop2ii}). Thus, in independent equilibria, both the number and identity of network contributors depend on the linking cost. This pattern is important for the results on how welfare and polarization change in the linking costs presented in Section \ref{largesocieties}.}

A difficulty in the analysis is that the realization of players' types determines which equilibria are possible. \added{Additionally, equilibrium multiplicity makes comparative statics non-straightforward. To obtain robust results, we turn to the study of societies with infinitely many players and types.} Following \cite{galeotti2010law}, we define a \textit{large society} as an economy where $n\to\infty$, and we impose that the resulting type distribution $\Tau_{\infty}$ is strictly increasing for any type $t\in(0,1)$, so that the density of players is non-zero everywhere. In words, there is an infinite number of players and the realization of types is ``dense'', \added{reflecting settings like social media or the internet, where virtually every preference and type of content can be found.}
\begin{proposition}\label{prop:lotf}
In any Nash equilibrium $\mathbf{s}^\ast$ of a large society \added{where $n\to\infty$ and $\Tau_{\infty}$ is strictly increasing in $t\in(0,1)$}, $\vert C(\mathbf{g}^\ast)\vert/n\to0$.
\end{proposition}
\added{This is the law of the few: as the number of players grows, only a vanishing fraction of them are network contributors.} In collaborative equilibria, this is immediate, as there are always exactly two core contributors. In independent equilibria, each network contributor receives links from players with tastes in an interval around their own. \added{If there were too many contributors, these intervals would necessarily overlap, which would create a profitable deviation for at least one contributor and therefore rule out such a configuration.} This bounds the set of potential network contributors.\footnote{In our model, the law of the few arises for different reasons than in single-public-good models \citep{galeotti2010law}. Here, it emerges because each network contributor has a group of free riders with similar tastes.}

\added{This result captures a stylized fact of online platforms such as Facebook, Twitter (now X) and TikTok: follower and activity distributions are highly skewed, with a small fraction of users producing most content while the majority primarily consume \citep{shafiq2013identifying,twitter,tiktok2024us}.}

\added{While by equilibrium multiplicity the number of network contributors can change non-monotonically,} the law of the few identifies when each class of equilibria exists.
\begin{proposition}\label{prop:eq_net}
In a large society there exists a \added{threshold} $K$ such that, if $k>K$, the unique equilibrium network is empty, \added{whereas for $k\le K$ an independent equilibrium exists.} 
If Assumption~\ref{ass:zeal} holds, there also exist thresholds $\underline{k}_{C}$, $\bar{k}_{C}$, and $z_{C}$ such that a collaborative equilibrium exists if $z(f,c)\le z_{C}$ and $\underline{k}_{C}\le k\le \bar{k}_{C}$.
\end{proposition}
If the value of the linking cost is too high, the only equilibrium network is empty. Otherwise, an independent equilibrium always exists.

For links among network contributors to be profitable, they need to be somewhat moderate. However, the existence of moderate network contributors requires that, \textit{(i)}, linking costs are sufficiently low that linking to moderate types is profitable, \textit{(ii)}, linking costs are sufficiently high that linking to extreme types is not, and, \textit{(iii)}, moderate types' total demand of the public goods is high enough compared to players with extreme tastes, i.e., if zeal is low enough. \added{Hence, the low polarization characteristic of collaborative equilibria arises only when these conditions are met; for a deeper analysis of polarization and welfare implications, we next turn to welfare-maximizing Nash equilibrium.}

\section{The Welfare-Maximizing Nash Equilibrium}\label{largesocieties}

The identity of the network contributors influences how much other players free ride by linking to them. This, in turn, affects both overall welfare and the polarization of consumption bundles. In this section, we examine how linking costs and preferences shape these outcomes.

We focus on the welfare-maximizing equilibrium\added{, since changes in linking costs can alter which equilibria exist in a non-monotonic way, preventing simple comparative statics across all equilibria}. Considering the equilibrium that maximizes welfare is a standard approach \citep[e.g.,][]{bramoulle2007public}. Moreover, since our main finding is that lower linking costs can reduce welfare, it makes sense to restrict attention to the ``best" Nash equilibrium.\footnote{The efficient outcome would be a star network, where the hub contributes more public goods than in isolation. However, this is not an equilibrium (see Section \ref{efficient}).}

\added{We first give sufficient conditions identifying the welfare-maximizing equilibrium and its dependence on zeal and linking costs.}\footnote{\added{As we show in the Supplemental Appendix, the welfare-minimizing equilibrium has a similar structure, so the comparative statics results on linking costs presented below still apply. The key difference lies in that in the welfare-minimizing equilibrium, both network contributors link to each other and produce relatively little, which explains why this configuration minimizes welfare.}}
\begin{proposition}\label{prop:welfare-max}
If Assumption~\ref{ass:zeal} holds, then in a large society there exist thresholds 
$\underline{z}^W_C$, $\bar{z}^W_{C}$, $\underline{k}^W_{C}$, and $\bar{k}^W_{C}$ such that \added{the 
welfare-maximizing equilibrium is:}
\begin{enumerate}[{(}i{)}]
    \item \added{a star if $z(f,c)\le \underline{z}^W_C$;}
    \item \added{collaborative if $\underline{z}^W_C < z(f,c)\le \bar{z}^W_C$ and $k\in[\underline{k}^W_C,\bar{k}^W_C]$, and independent otherwise;}
    \item \added{independent if $z(f,c)>\bar{z}^W_C$.}
\end{enumerate}
\end{proposition}
\added{In the welfare-maximizing equilibrium, network contributors do not mutually link to each other, since that would excessively reduce their provision. When zeal is high, a few extreme contributors supply large total bundles, and concentrating links on them increases overall consumption. When zeal is low or negative, moderate contributors attract links from nearby types, giving more players access to balanced bundles.}

\added{Collaborative equilibria maximize welfare only when exactly one network contributor links to the other and additional conditions are satisfied. A collaborative equilibrium must exist (Proposition \ref{prop:eq_net}) and zeal must be intermediate: too low, and independent equilibria with moderate network contributors emerge; too high, and extreme contributors provide so much that a moderate collaborative contributor is unable to match their total provision. More generally, the welfare-maximizing outcome reflects the interaction between zeal and linking costs.}

\subsection{Welfare and Polarization}

We now examine how changes in linking costs affect welfare and polarization. Intuitively, one might expect higher linking costs to always reduce welfare and increase polarization. As linking costs decrease, more players are likely to free ride, which should boost their utility and lead to more similar consumption bundles, thus reducing polarization. However, the next proposition shows that higher linking costs can sometimes increase the number of network contributors, thereby improving welfare. The proposition outlines the conditions under which this occurs \added{focusing on situations in which the welfare-maximizing equilibrium is independent}. Denote by $\mathbf{s}^W(k)$ the welfare-maximizing equilibrium when the linking cost is $k$. Then:
\begin{proposition}\label{prop:welfare}
If Assumption~\ref{ass:zeal} holds, then in a large society there exist thresholds 
$\underline{z}$, $\bar{z}$, $\bar{k}_2$, $\bar{k}_3$, and $\Delta>0$ such that, 
for $k$ with $\bar{k}_2<k+\Delta<\bar{k}_3$ and intermediate zeal 
$z(f,c)\in[\underline{z},\bar{z}]$, welfare is higher in 
$\mathbf{s}^W(k+\Delta)$ \added{than in} $\mathbf{s}^W(k)$.
\end{proposition}
An increase in linking costs can alter the set of network contributors, potentially raising welfare if a moderate network contributor with sufficiently large public good provision emerges. When linking costs are low enough ($k<\bar{k}_2$), a moderate contributor links to players with extreme tastes. In this case, the welfare-maximizing equilibrium includes two network contributors with extreme taste. However, when linking costs increase to $k\in(\bar{k}_2,\bar{k}_3)$, moderate players no longer wish to free ride on extreme-taste players, and a moderate network contributor emerges. Specifically, if zeal is intermediate, the set of contributors expands to include a moderate player alongside two extreme-taste players, since the moderate player's demand for one public good is not high enough to attract links from them.

The presence of a network contributor with moderate taste benefits players of sufficiently similar types. The lower the zeal, the broader the range of types that gain more from linking to this moderate contributor. Therefore, for any distribution of types, there exists a sufficiently low level of zeal that ensures a critical mass of agents experiences increased utility following the rise in linking costs.\footnote{With negative zeal and low initial linking costs, increasing linking costs can improve welfare if network contributors have closer tastes, making them more moderate. Indeed, since zeal is negative, more moderate contributors provide a greater total amount of public goods.}

The identity of the network contributors determines free riders' consumption bundles, and hence the level of polarization. We now show that polarization in the welfare-maximizing equilibrium varies non-monotonically with linking costs.
\begin{proposition}\label{prop:polarization}
Suppose Assumption~\ref{ass:zeal} holds and zeal is \added{nonzero}. Then, in a large society:
\begin{enumerate}[{(}i{)}]
    \item There exists a threshold $\underline{k}$ such that polarization in the welfare-maximizing equilibrium increases with the linking cost for $k\in(0,\underline{k}]$. 
    \item If zeal is strictly positive, there exist thresholds $\bar{k}$ and $\xi$ such that polarization decreases with the linking cost for $z(f,c)<\xi$ and $k\in[\bar{k},K]$.
\end{enumerate}
\end{proposition}
If the linking cost is sufficiently low, there are always two network contributors in the welfare-maximizing equilibrium, with some players linking to both. In this scenario, an increase in the linking cost reduces the number of players who link to both network contributors. As fewer players consume the same public good bundle, public good consumption becomes more polarized.

If zeal is positive, the set of network contributors in the welfare-maximizing equilibrium includes two players with extreme tastes when the linking cost is high. Players with moderate tastes are isolated, and rising linking costs expand this set. Since isolated players consume moderate bundles (their isolation demands) rather than extreme bundles from contributors, the consumption bundles become less polarized, leading to a decrease in polarization. However, this effect requires zeal to be sufficiently low: if zeal is too high, extreme contributors produce so much that even moderate players find it profitable to follow them despite high linking costs.

When zeal is negative and linking costs are high, there is only one network contributor, who is moderate. Therefore, in this scenario, the effect of an increase in linking costs on polarization depends on the distribution of types.

Finally, when zeal is zero, all players demand the same total amount of public good, which shuts down the mechanism behind the results of Proposition \ref{prop:polarization}.

\begin{figure}[htbp!]
\begin{center}
    \begin{subfigure}{0.4\linewidth}
    \resizebox{5cm}{!}{
    \begin{tikzpicture}[
	contributor/.style={diamond,draw,thick, fill=blue!20, scale=0.7},
	periphery/.style={circle,draw,thick, scale=0.7, fill=yellow!20},
	core/.style={regular polygon,regular polygon sides=5,draw,thick, fill=green!20, scale=0.7},
	]
	\path[draw,thick,below] (0,0) edge node {} (5,0);
	\path[draw,thick,left] (0,0) edge node {} (0,4);
	\node[contributor] at (5,3.5) (1){1};
	\node[diamond,draw,thick, fill=red!20, scale=0.7] at (0,3.5) (0){0};
    \draw[->,thin] (2.05,0.2)--(0);
    \draw[->,thin] (2.95,0.2)--(1);
	\fill[blue,nearly transparent] (5,0.2) -- (2.95,0.2) -- (2.95,-0.2) -- (5,-0.2) -- cycle;
	\fill[pattern = north east lines] (5,0.2) -- (2.95,0.2) -- (2.95,-0.2) -- (5,-0.2) -- cycle;
	\fill[red,nearly transparent] (0,0.2) -- (2.05,0.2) -- (2.05,-0.2) -- (0,-0.2) -- cycle;
	\fill[pattern = crosshatch dots] (0,0.2) -- (2.05,0.2) -- (2.05,-0.2) -- (0,-0.2) -- cycle;
	\end{tikzpicture}}
	\caption{High linking cost: \\ $k=.73$;\\ $W(\mathbf{s}^{W(.73)})=4.3$;\\ $\rho(\mathbf{s}^{W(.73)})=358273$}
	\end{subfigure}
    \begin{subfigure}{0.4\linewidth}
    \resizebox{5cm}{!}{
    \begin{tikzpicture}[
	contributor/.style={diamond,draw,thick, fill=blue!20, scale=0.7},
	periphery/.style={circle,draw,thick, scale=0.7, fill=yellow!20},
	core/.style={regular polygon,regular polygon sides=5,draw,thick, fill=green!20, scale=0.7},
	]
	\path[draw,thick,below] (0,0) edge node {} (5,0);
	\path[draw,thick,left] (0,0) edge node {} (0,4);
	\node[contributor] at (5,3.5) (1){1};
	\node[diamond,draw,thick, fill=red!20, scale=0.7] at (0,3.5) (0){0};
	\node[diamond,draw,thick, fill=green!20, scale=0.7] at (2.5,2.8) (m){m};
	\draw[->,thin] (2.87879,0.2)--(1);
	\draw[->,thin] (2.12121,0.2)--(0);
 	\draw[->,thin] (2.87879,0.2)--(m);
	\draw[->,thin] (2.12121,0.2)--(m);
	\fill[blue,nearly transparent] (5,0.2) -- (2.87879,0.2) -- (2.87879,-0.2) -- (5,-0.2) -- cycle;
	\fill[pattern = north east lines] (5,0.2) -- (2.87879,0.2) -- (2.87879,-0.2) -- (5,-0.2) -- cycle;
	\fill[red,nearly transparent] (0,0.2) -- (2.12121,0.2) -- (2.12121,-0.2) -- (0,-0.2) -- cycle;
	\fill[pattern = crosshatch dots] (0,0.2) -- (2.12121,0.2) -- (2.12121,-0.2) -- (0,-0.2) -- cycle;
	\fill[green, nearly transparent] (2.12121,0.2) -- (2.87879,0.2) -- (2.87879,-0.2) -- (2.12121,-0.2) -- cycle;
	\fill[pattern = vertical lines] (2.12121,0.2) -- (2.87879,0.2) -- (2.87879,-0.2) -- (2.12121,-0.2) -- cycle;
	\end{tikzpicture}}
	\caption{Intermediate linking costs: \\
 $k=.677$ \\ $W(\mathbf{s}^{.677)})=4.35$\\ $\rho(\mathbf{s}^{W(.677)})=19923$}
	\end{subfigure}
\begin{subfigure}{0.4\linewidth}
\resizebox{5cm}{!}{
    \begin{tikzpicture}[
	contributor/.style={diamond,draw,thick, fill=blue!20, scale=0.7},
	transparentperiphery/.style={circle,draw,thick, scale=0.7, fill=white, transparent},
	periphery/.style={circle,draw,thick, scale=0.7, fill=yellow!20},
	core/.style={regular polygon,regular polygon sides=5,draw,thick, fill=green!20, scale=0.7},
	]
	\path[draw,thick,below] (0,0) edge node {} (5,0);
	\path[draw,thick,left] (0,0) edge node {} (0,4);
	\node[periphery] at (5,1.6) (1){1};
	\node[periphery] at (0,1.6) (0){0};
	\node[contributor] at (4.45,3.2) (A){b};
	\node[diamond,draw,thick, fill=red!20, scale=0.7] at (0.55,3.2) (B){a};
	\draw[->,thin] (2.5,0)--(A);
	\draw[->,thin] (2.5,0)--(B);
	\draw[->,thin] (1)--(A);
	\draw[->,thin] (0)--(B);
	\fill[blue,nearly transparent] (5,0.2) -- (2.5,0.2) -- (2.5,-0.2) -- (5,-0.2) -- cycle;
	\fill[pattern = north east lines] (5,0.2) -- (2.5,0.2) -- (2.5,-0.2) -- (5,-0.2) -- cycle;
	\fill[red,nearly transparent] (0,0.2) -- (2.5,0.2) -- (2.5,-0.2) -- (0,-0.2) -- cycle;
	\fill[pattern = crosshatch dots] (0,0.2) -- (2.5,0.2) -- (2.5,-0.2) -- (0,-0.2) -- cycle;
	\end{tikzpicture}
    }
	\caption{Low linking costs:\\ $k=.676$ \\
    $W(\mathbf{s}^{W(.676)})=4.32$ \\
    $\rho(\mathbf{s}^{W(.676)})=51266$}
	\end{subfigure}
\begin{subfigure}{0.4\linewidth}
\resizebox{5cm}{!}{
    \begin{tikzpicture}[
	contributor/.style={diamond,draw,thick, fill=blue!20, scale=0.7},
	transparentperiphery/.style={circle,draw,thick, scale=0.7, fill=white, transparent},
	periphery/.style={circle,draw,thick, scale=0.7, fill=yellow!20},
	core/.style={regular polygon,regular polygon sides=5,draw,thick, fill=green!20, scale=0.7},
	]
	\path[draw,thick,below] (0,0) edge node {} (5,0);
	\path[draw,thick,left] (0,0) edge node {} (0,4);
	\node[contributor] at (5,3.5) (1){1};
	\node[diamond,draw,thick, fill=red!20, scale=0.7] at (0,3.5) (0){0};
	\draw[->,thin] (2.5,0.2)--(1);
	\draw[->,thin] (2.5,0.2)--(0);
	\draw[->,thin] (4.75,0.2)--(1);
    \draw[->,thin] (.25,0.2)--(0);
	\fill[blue,nearly transparent] (5,0.2) -- (4.75,0.2) -- (4.75,-0.2) -- (5,-0.2) -- cycle;
	\fill[pattern = north east lines] (5,0.2) -- (4.75,0.2) -- (4.75,-0.2) -- (5,-0.2) -- cycle;
	\fill[red,nearly transparent] (0,0.2) -- (.25,0.2) -- (.25,-0.2) -- (0,-0.2) -- cycle;
	\fill[pattern = crosshatch dots] (0,0.2) -- (.25,0.2) -- (.25,-0.2) -- (0,-0.2) -- cycle;
	\fill[green, nearly transparent] (.25,0.2) -- (4.75,0.2) -- (4.75,-0.2) -- (.25,-0.2) -- cycle;
	\fill[pattern = vertical lines] (.25,0.2) -- (4.75,0.2) -- (4.75,-0.2) -- (.25,-0.2) -- cycle;
	\end{tikzpicture}
    }
	\caption{Very low linking costs:\\ $k=.14$ \\
    $W(\mathbf{s}^{W(.14)})=12.2$ \\
    $\rho(\mathbf{s}^{W(.14)})=2671$
    }
	\end{subfigure}
\end{center}
\caption{Welfare-maximizing equilibrium for different values of the linking cost, $k$.}\label{fig:welfare-pol-concave}
\smallskip
\footnotesize
\emph{Notes:} In the example, $f(\cdot)=(\cdot)^{.15}$ and $c=10^{-5}$ and the type distribution is a Normal distribution truncated at $0$ and $1$ with mean $.5$ and variance $1$. The horizontal axis represents types from $0$ to $1$, and the vertical axis represents players' total public good contributions. Arrows represent links, the shaded colored area in the horizontal axis represents neighborhoods (white color represents isolated players). If $k=.82$, the welfare-maximizing equilibrium is empty, $W(\mathbf{s}^{W(.82)})=4.25$ and $\rho(\mathbf{s}^{W(.82)})=13725$.
\end{figure}

Figure \ref{fig:welfare-pol-concave} illustrates the results of Propositions \ref{prop:welfare} and \ref{prop:polarization} when zeal is positive. When linking costs are high (\(k=.73\), Panel a), some players are isolated, welfare is low, and polarization is high as free riders access extreme bundles. Slightly lower linking costs increase welfare, as more players find it optimal to free ride, but it also increases polarization as more players consume a large amount of one good.

When linking costs reach $.677$ (Panel (b)), a moderate contributor, $m$, appears. As zeal is sufficiently low, players with similar taste gain enough to increases welfare. Polarization decreases as many free riders now consume $m$'s moderate bundle.

At $k=.676$ (Panel (c)), $m$ is no longer a network contributor in the welfare-maximizing equilibrium. Free riders near $m$ now link to contributors with more extreme tastes, accessing bundles that are less aligned with their preferences. As a result, their utility declines, aggregate welfare falls, and polarization rises.

When linking costs are low ($k=.14$, Panel (d)), many players free ride on both network contributors, so they consume nearly identical bundles. Polarization is then low. Moreover, welfare is high since players can access information at a low cost. Further cost reductions amplify these effects as more players consume similar bundles.

\added{The key insight is that even in welfare-maximizing equilibria, lowering linking costs—through platform design choices such as recommendation algorithms, frictionless following, or interface adjustments—can have non-monotonic effects on both welfare and polarization. These effects arise from the trade-off between preferring content produced by like-minded users and content produced by highly active users, which determines who becomes a network contributor. Understanding these equilibrium responses is crucial for evaluating policy interventions analyzed below.}

\subsection{Efficient Solution}\label{efficient}

We first derive \added{the allocation that maximizes total welfare, irrespective of whether it constitutes a Nash equilibrium}. If linking costs are not too high, the efficient solution forms a star network, as this minimizes linking costs.\footnote{We omit the proof as it follows similar arguments of \cite{kinateder2017public}.}
\begin{corollary}
    If the linking cost is sufficiently low, the efficient solution is a star network, with potentially some isolated players.
\end{corollary}
Several points are worth noting. The hub's provision of public goods is designed to internalize the benefits for all players in the star. Consequently, the planner may dictate a significantly larger provision than what occurs in a decentralized equilibrium, meaning the efficient network does not constitute an equilibrium. As a result, the efficient network can still may be a star even when linking costs are such that the only Nash equilibrium is an empty network.

In the next section, we explore how a social planner can intervene to bring welfare in the Nash equilibria of the game closer to that of the efficient solution. 

\subsection{Policy Interventions}

\paragraph{Subsidies to Provision.} The first policy intervention we study involves providing \added{incentives to certain users to increase their contributions to public goods. Examples include rewards for creating high-quality content on platforms such as YouTube or TikTok, or incentives for sharing reliable information in online communities.}

The central question is who these policies should target. For instance, \added{rewards might target creators who produce content that appeals to broad audiences, or users who provide cross-cutting perspectives that connect otherwise separated communities. The impact of such policies depends on which users are subsidized, shaping both the diversity and total amount of information circulating on the platform.}

Formally, suppose the social planner has a budget $V$ to subsidize public good provision. Denote by $\mathbf{v}$ the vector whose entry $v_i$ denotes the subsidy to player $i$ such that $c_i=\max\{c-v_i,0\}$ and the total subsidy player $i$ receives is $v_i(x_i+y_i)$. The isolation demand of a player $i$ who is subsidized is then $\hat{x}^{v}_i=f^{'-1}(\frac{c-v_i}{t_i})$ and $\hat{y}^v_i=f^{'-1}(\frac{c-v_i}{1-t_i})$. Denote by $\mathbf{s}^{W(\mathbf{v})}$ the welfare-maximizing equilibrium given a subsidy vector $\mathbf{v}$, so that $\mathbf{s}^{W(0)}=\mathbf{s}^W$. The problem for the policy maker is:
\begin{eqnarray*}
\max_{\mathbf{v}} \sum_{i\in N}U_i(\mathbf{s}^{W(\mathbf{v})})\\
\text{s.t. }\sum_{i\in N}v_i(x_i+y_i)\leq V. 
\end{eqnarray*}
We establish the following result.\footnote{\added{Some remarks are in order. First, while we focus on Nash implementation, multiple equilibria generally exist, and we do not analyze how subsidies affect the full set of equilibria. We also assume that the social planner has accurate knowledge of the type distribution. In the large‑budget regime, however, the equilibrium is unique, so the predictions of the subsidy intervention are unambiguous; for smaller budgets, outcomes may depend on the selected equilibrium and parameter estimates. Second, if subsidies were good‑specific rather than targeted to players, two possibilities arise with a sufficiently large budget: if zeal is low relative to linking costs, a single contributor emerges and forms a star, as under individual subsidies; if zeal is high, the two players with the most extreme tastes become contributors instead. As With smaller budgets, good‑specific subsidies are not optimal, as much of the subsidy goes to players who generate no externalities because no one links to them.}}
\begin{proposition}\label{prop:subsidy}
If Assumption~\ref{ass:zeal} holds, in a large society there exist thresholds $\underline{V}$ and $\overline{V}$ such that the welfare-maximizing subsidy allocation $\mathbf{v}$ leads to the following patterns in network contributors:
\begin{enumerate}[{(}i{)}]
    \item If $V < \underline{V}$, \added{only original network contributors may receive a positive subsidy,} i.e., $v_i>0$ if $i \in C(\mathbf{g}^W)$.
    \item If $V \in [\underline{V},\overline{V})$, \added{at least one player who was not a contributor before the subsidy becomes a contributor}, i.e., $v_m>0$ for some $m \not\in C(\mathbf{g}^W)$ and $\vert C(\mathbf{g}^{W(\mathbf{v})})\vert \ge 2$.
    \item If $V \ge \overline{V}$, \added{a single player becomes the sole network contributor after the subsidy,} i.e., $C(\mathbf{g}^{W(\mathbf{v})}) = \{m\}$ and $g^W_{im} = 1$ for all $i \in N \setminus \{m\}$.
\end{enumerate}
\end{proposition}
First, players receive a subsidy only if they have links in the resulting welfare-maximizing equilibrium, as subsidies are effective if they generate more spillovers for free riders. Second, subsidies increase a player's demand for public goods, making free riding more advantageous. However, with a limited budget, the social planner cannot make it profitable to link to a player who would otherwise free ride. Therefore, the social planner focuses on subsidizing existing network contributors.

With a larger budget, subsidies can target players who would not receive links without the subsidy, particularly if they are situated in densely populated areas of the type space, as this fosters network spillovers. If the subsidy is sufficiently large, this player becomes the sole recipient of links, resulting in a star network. 

However, the hub of this star does not provide the same bundle as in the efficient solution. Instead, the hub will offer a public goods bundle based on its own preferences, without considering the preferences of the free riders relying on it. \added{In contrast to \cite{kinateder2023free}, where a planner always induces a star with a single public good, in our setting a star emerges only with a sufficiently large budget.}

\begin{figure}[ht]
    \centering
    \begin{subfigure}[b]{0.32\linewidth}
    \centering
    \begin{tikzpicture}[
	contributor/.style={diamond,draw,thick, fill=blue!20, scale=0.7},
	periphery/.style={circle,draw,thick, scale=0.7, fill=yellow!20},
	transparentperiphery/.style={circle,draw,thick, scale=0.7, fill=white, transparent},
	core/.style={regular polygon,regular polygon sides=5,draw,thick, fill=green!20, scale=0.7},
	isolated/.style={regular polygon,regular polygon sides=7,draw,thick, fill=red!20, scale=0.7},
	]
	\path[draw,thick,below] (0,0) edge node {} (4.2,0);
	\path[draw,thick,left] (0,0) edge node {} (0,3.5);
	\node[contributor] at (3.53,1) (i){i};
	\node[contributor] at (.67,1) (j){j};
	\draw[->,thin] (4.2,0)--(i);
	\draw[->,thin] (0,0)--(j);
	\draw[->,thin] (2.1,0)--(i);
	\draw[->,thin] (2.1,0)--(j); [black,midway,xshift=9pt,anchor = north] {\footnotesize $I(\mathbf{g^\ast})$};
	\fill[blue,nearly transparent] (4.2,0.2) -- (2.9,0.2) -- (2.9,-0.2) -- (4.2,-0.2) -- cycle;
	\fill[pattern = north east lines] (4.2,0.2) -- (2.9,0.2) -- (2.9,-0.2) -- (4.2,-0.2) -- cycle;
	\fill[red,nearly transparent] (0,0.2) -- (1.3,0.2) -- (1.3,-0.2) -- (0,-0.2) -- cycle;
	\fill[pattern = crosshatch dots] (0,0.2) -- (1.3,0.2) -- (1.3,-0.2) -- (0,-0.2) -- cycle;
	\fill[green, nearly transparent] (1.3,0.2) -- (2.9,0.2) -- (2.9,-0.2) -- (1.3,-0.2) -- cycle;
	\fill[pattern = vertical lines] (1.3,0.2) -- (2.9,0.2) -- (2.9,-0.2) -- (1.3,-0.2) -- cycle;
	\end{tikzpicture}
	\caption{$V\in[0,.546)$}
	\end{subfigure}
    \begin{subfigure}[b]{0.32\linewidth}
    \centering
    \begin{tikzpicture}[
	contributor/.style={diamond,draw,thick, fill=blue!20, scale=0.7},
	periphery/.style={circle,draw,thick, scale=0.7, fill=yellow!20},
	transparentperiphery/.style={circle,draw,thick, scale=0.7, fill=white, transparent},
	core/.style={regular polygon,regular polygon sides=5,draw,thick, fill=green!20, scale=0.7},
	isolated/.style={regular polygon,regular polygon sides=7,draw,thick, fill=red!20, scale=0.7},
	]
	\path[draw,thick,below] (0,0) edge node {} (4.2,0);
	\path[draw,thick,left] (0,0) edge node {} (0,4);
	\node[contributor] at (4.2,1) (1){1};
	\node[contributor] at (0,1) (0){0};
	\node[contributor] at (2.1,.84) (m){m};
	\draw[->,thin] (m)--(1);
	\draw[->,thin] (m)--(0);
	\draw[->,thin] (2.9,0)--(1);
	\draw[->,thin] (1.3,0)--(0);
	\draw[->,thin] (2.1,0)--(m);
	\fill[blue,nearly transparent] (4.2,0.2) -- (2.15,0.2) -- (2.15,-0.2) -- (4.2,-0.2) -- cycle;
	\fill[pattern = north east lines] (4.2,0.2) -- (2.15,0.2) -- (2.15,-0.2) -- (4.2,-0.2) -- cycle;
	\fill[red,nearly transparent] (0,0.2) -- (2.05,0.2) -- (2.05,-0.2) -- (0,-0.2) -- cycle;
	\fill[pattern = crosshatch dots] (0,0.2) -- (2.05,0.2) -- (2.05,-0.2) -- (0,-0.2) -- cycle;
	\fill[green, nearly transparent] (2.05,0.2) -- (2.15,0.2) -- (2.15,-0.2) -- (2.05,-0.2) -- cycle;
	\fill[pattern = vertical lines] (2.05,0.2) -- (2.15,0.2) -- (2.15,-0.2) -- (2.05,-0.2) -- cycle;
	\end{tikzpicture}
	\caption{$V\in[.546,2.686)$}
	\end{subfigure}\begin{subfigure}[b]{0.32\linewidth}
    \centering
    \begin{tikzpicture}[
	contributor/.style={diamond,draw,thick, fill=blue!20, scale=0.7},
	periphery/.style={circle,draw,thick, scale=0.7, fill=yellow!20},
	transparentperiphery/.style={circle,draw,thick, scale=0.7, fill=white, transparent},
	core/.style={regular polygon,regular polygon sides=5,draw,thick, fill=green!20, scale=0.7},
	isolated/.style={regular polygon,regular polygon sides=7,draw,thick, fill=red!20, scale=0.7},
	]
	\path[draw,thick,below] (0,0) edge node {} (4.2,0);
	\path[draw,thick,left] (0,0) edge node {} (0,4);
	\node[contributor] at (2.1,3.7) (m){m};
	\draw[->,thin] (4.2,0)--(m);
	\draw[->,thin] (0,0)--(m);
	\draw[->,thin] (2.9,0)--(m);
	\draw[->,thin] (1.3,0)--(m);
	\fill[green, nearly transparent] (0,0.2) -- (4.2,0.2) -- (4.2,-0.2) -- (0,-0.2) -- cycle;
	\fill[pattern = vertical lines] (0,0.2) -- (4.2,0.2) -- (4.2,-0.2) -- (0,-0.2) -- cycle;
	\end{tikzpicture}
	\caption{$V\geq2.868$}
	\end{subfigure}\\
    \begin{tikzpicture}[
	contributor/.style={diamond,draw,thick, fill=blue!20, scale=0.7},
	periphery/.style={circle,draw,thick, scale=0.7, fill=yellow!20},
	core/.style={regular polygon,regular polygon sides=5,draw,thick, fill=green!20, scale=0.7},
	isolated/.style={regular polygon,regular polygon sides=7,draw,thick, fill=red!20, scale=0.7},
	]
	\centering
	\matrix [draw,below left] at (current bounding box.north west) {
  \node [contributor,label=right:Contributor] {}; &
  \node [periphery,label=right:Periphery] {}; &
  \node [isolated,label=right:Isolated] {};\\
};
    \end{tikzpicture}
\caption{Examples of independent equilibria.}\label{fig-prop8}
\begin{minipage}{\linewidth}
\smallskip
\footnotesize
\emph{Notes:} In this example, $f(\cdot)=\ln(\cdot)$, $c=1$, and $k=.84$. The type distribution is a Normal distribution truncated at $0$ and $1$ with mean $.5$ and variance $1$. The horizontal axis represents types ranging from $0$ to $1$, and the vertical axis represents total public good contributions of players. Arrows represent links.
\end{minipage}
\end{figure}

Figure \ref{fig-prop8} illustrates an example of Proposition \ref{prop:subsidy}. 
In the welfare-maximizing equilibrium without subsidies, players $i=0.84$ and $j=0.16$ are the network contributors. The moderate player $m=0.5$ receives links if and only if $c\big(x_m^{W(\mathbf{v})}+y_m^{W(\mathbf{v})}\big)\ge k=0.84$. This requires a subsidy $v_m=0.65$, which raises $m$’s isolation demand to $\hat{x}_m^{v_m} = \hat{y}_m^{v_m} = 1.43$, partly met by linking to players $0$ and $1$. Unlike the no-subsidy case, $m$ can then attract links while still free riding provided the total budget satisfies $V\ge \underline{V}=2\times(0.65\cdot 0.84/2)$, where $x_m^{W(\mathbf{v})}=y_m^{W(\mathbf{v})}=0.84/2$.

Players $1$ and $0$ are now also network contributors and $m$ free rides on them. To induce a star, $0$ and $1$ must also free ride on $m$. This requires $m$'s contributions to $x$ and $y$ to be sufficiently large to receive links from all types. Specifically, this occurs when: $c x_m^{v_m} \geq k = 0.84 \text{ and } c y_m^{v_m} \geq k = 0.84$. The necessary subsidy is $v_m = 0.73$, which implies the following isolation demands: $\hat{x}_m^{v_m} = \hat{y}_m^{v_m} = 1.84$. Consequently, the total budget required to induce the star is: $\bar{V} = 1.84 \cdot 0.73 + 1.84 \cdot 0.73 = 2.686$.

\added{Proposition \ref{prop:subsidy} shows that the optimal allocation of rewards depends on the size of the budget. With modest rewards, it is most effective to support highly active creators who already generate significant spillovers. With larger budgets, it becomes more beneficial to subsidize creators whose content reaches broader or more diverse audience---even if they aren’t initially the most prolific. This mirrors real‐world patterns on major platforms: for example, TikTok's revamped Creator Rewards Program rewards original, high-quality videos of at least one minute, using metrics like engagement, play duration, and originality rather than sheer volume \citep{tiktok2025}. Meanwhile, platforms like YouTube monetize creators based on watch time and engagement, which can indirectly favor longer or more engaging content, rather than simply rewarding rapid volume generation \citep{youtube}. Proposition \ref{prop:subsidy} formalizes this trade-off between reinforcing existing creator activity and fostering broader impact across the content ecosystem.}

\paragraph{Subsidies to Links.} \added{A social planner may also incentivize information diffusion by subsidizing the formation of links rather than contributions. In the context of social media, this can be interpreted as lowering the effective cost of following certain users, for example through platform algorithms that make some accounts easier to find or recommend, capturing the role of algorithmic amplification in shaping the network.}

\added{Formally, let $l_{ji} \geq 0$ be the subsidy to player $j$ for linking to player $i$, so that the effective linking cost is $k - l_{ji}$. Let $L(\mathbf{l})$ denote the total subsidy expenditure under scheme $\mathbf{l}$, so that $L(\mathbf{l}) = \sum_{j,i} l_{ji} \cdot g_{ji}^{W(\mathbf{l})}$, where $g_{ji}^{W(\mathbf{l})}$ is the equilibrium link decision. The social planner faces a budget constraint $L(\mathbf{l}) \le \bar{L}$. Let $\mathbf{s}^{W(\mathbf{l})}$ denote the welfare-maximizing equilibrium under the link-subsidy scheme $\mathbf{l}$.}
\begin{corollary}\label{cor:linksubsidy}
\added{Suppose Assumption~\ref{ass:zeal} holds. 
If a link-subsidy scheme $\mathbf{l}$ does not change the set of network contributors, i.e.,
$C(\mathbf{g}^{W(\mathbf{l})})=C(\mathbf{g}^{W(\mathbf{0})})$,
then the welfare gain induced by the subsidies is at most $L(\mathbf{l})$.}
\end{corollary}
\noindent \added{Notably, link subsidies leave content provision unchanged. Moreover, when they do not alter the set of network contributors, they may induce additional links and hence additional spillovers, but such links are formed only when the associated net gains equal the subsidized linking cost. Because linking costs enter utility linearly, the resulting welfare gain is therefore bounded by the budget used. Additional welfare gains require subsidies large enough to reshape which users become network contributors.}

\added{This bound highlights a marginal tradeoff between subsidizing content provision and subsidizing links. With decreasing marginal returns to content, additional production generates large welfare gains when aggregate provision is low, but increasingly small gains when users already consume substantial amounts of public goods. By contrast, link subsidies operate through a linear reduction in linking costs and thus deliver approximately constant marginal gains once no new links are induced. As a result, reallocating attention through links is most effective only in environments where content is already abundant and marginal utility from further provision is low.}

\added{This logic is illustrated by Proposition\ref{prop:subsidy}.(iii), in which sufficiently large provision subsidies generate a star network with a single hub. In this case, welfare is already high and, because all users are exposed to the same source, polarization is mechanically low. Link subsidies are therefore naturally interpreted as a secondary instrument, relevant once such high-provision network structures are already in place.}

\added{The key message is that policies that reallocate attention—through recommendation algorithms or interface design that makes some users easier to follow—primarily reshuffle exposure rather than change who produces information. In the context of social media, reducing frictions in following or amplifying existing creators improves welfare mainly through mechanical cost savings. This also helps explain why exposure-based interventions can have ambiguous effects on polarization: unless such policies alter incentives to produce content, they change consumption patterns without shifting the underlying distribution of information.}

\section{Discussion}\label{discussion}

We now discuss some extensions to the benchmark model. Proofs and additional results appear in the Supplemental Appendix, which also examines intrinsic motivation, information decay, complementarities between neighbors’ provisions, indirect spillovers, and heterogeneity in contribution and linking costs.

\bigskip

\textbf{Benefits from receiving links.} So far, we have not considered the possibility of benefits from incoming links. However, it is reasonable to assume that, in many cases, receiving links enhances status and, thus, payoffs. For instance, social media influencers appreciate having many followers. Here, we illustrate that our results are robust to such considerations.

We conceptualize the benefits from incoming links as \textit{status rents} as in \cite{van2020competition}. Formally, the utility of player $i$ is given by
\begin{equation}\label{eq:status}
    U_i(\mathbf{s})=t_if(x_i+\bar{x}_i(\mathbf{g}))+(1-t_i)f(y_i+\bar{y}_i(\mathbf{g}))-(x_i+y_i)c-\eta_i(\mathbf{g})k+\omega_i(\mathbf{g})b.
\end{equation}
The additional term in \ref{eq:status} is \(\omega_i(g)\), counting player \(i\)’s incoming links—those sponsored by others to free ride on \(i\)’s provision. Each incoming link yields a benefit \(b\) to \(i\), with \(b < k\) to ensure linking alone does not create welfare.
\begin{corollary}\label{cor:status}
    The set of equilibria in the model with status rents coincides with the set of equilibria in the baseline model. 
\end{corollary}
Demands in isolation are unaffected by benefits from incoming links, so incentives to sponsor links under status rents are the same as in the baseline model. Thus, the equilibria in Proposition \ref{prop:charact} remain valid. Consider any equilibrium with incoming-link benefits: a player may (i) delete an outgoing link, (ii) adjust contribution to good \(x\), or (iii) adjust contribution to good \(y\). By Lemma \ref{lemma:BK}, players without incoming links do not increase their contributions, since in their best reply they take others’ strategies as given. While players with incoming links receive weakly higher benefits than in the baseline, as they had no profitable deviation there, they have none here either.

\added{The static Nash framework eliminates strategic incentives to produce content for followers. This means polarization emerges from preference heterogeneity alone, establishing a baseline that platform engagement incentives—such as monetization based on followers that potentially generate superstar contributors, as in \citet{van2020competition}—would likely worsen while preserving equilibrium structures.}

\bigskip

\noindent \textbf{Two-way flow of spillovers.} In the baseline model, we assume a one-way flow of spillovers, i.e., to access the public good provision of a player, one has to link to that player. This assumption suits social media, where being exposed to a user's content requires linking to them. In other applications, two-way flow of spillovers, where the recipient of a link also accesses the linker's public good provision, is more appropriate \citep{galeotti2010law,kinateder2017public}.

Under two-way spillovers, players with extreme preferences ($0$ and $1$) are always the largest contributors to one good. This precludes a collaborative equilibrium and prevents other players from contributing in an independent equilibrium. If they did, they would attract links from higher-demand players, which, by Lemma \ref{lemma:BK}, would reduce their provision, yielding a contradiction. Hence, no equilibrium exists in the two-way flow model when linking costs are intermediate.

\bigskip

\noindent \textbf{Heterogeneous costs.} We assume that all players incur the same costs of producing the public good. \added{If a player has a sufficiently low cost, she will become the main contributor, generating a star network. This outcome is analogous to the one described in Proposition \ref{prop:subsidy}, where introducing a sufficiently large subsidy creates heterogeneity in effective costs and leads to a star: in both cases, one player provides the bulk of the public goods.} See \cite{kinateder2017public} for a more detailed analysis.

\bigskip

\noindent \textbf{Weighted links.}
In our model, a player can choose to either establish a full link or not link to another player, reflecting the discrete nature of social and professional relationships. However, there are situations where the strength of a link can be adjusted. For instance, \added{social media users determine how much time to dedicate to reading certain content}. In this sense, links can be weighted.

Formally, a player pays \(\alpha k\) to link to \(j\) and access \(\alpha x_j\) and \(\alpha y_j\) \citep{kinateder2022local}. Any player valuing a good links to its largest contributor if profitable. \added{As players with more extreme taste for a good always demand more of it,} in equilibrium at most two network contributors exist. Hence, welfare declines as linking costs rise, provided the network is non-empty. Since free riding is cheaper than own provision, free riders consume more than in isolation. Thus, the non-monotonic effect of linking costs on polarization also holds under weighted links.

\bigskip

\bigskip

\section{Conclusions}\label{conclusion}

\added{We study a game in which social media users provide different kinds of content—modeled as local public goods—with heterogeneous tastes, and choose whose content to consume. These choices generate an equilibrium network of information spillovers shaped by two forces: users prefer sources aligned with their own interests, but also seek out prolific providers of particular types of content. Although the model is static, it highlights that the equilibrium distribution of content—the network of contributors and followers—shapes who consumes what. Understanding these patterns is essential for anticipating how polarization evolves over time.}

\added{By examining the welfare-maximizing equilibrium, we reveal novel insights regarding how linking costs affect welfare and polarization. Changes in these costs alter who contributes information and who benefits from others’ contributions. Specifically, intermediate linking costs can encourage the emergence of additional contributors, improving both welfare and diversity, while very low costs may discourage moderate contributors, concentrating exposure on extreme sources. The resulting relationship between linking costs and polarization is non-monotonic, implying that technologies designed to facilitate communication may paradoxically worsen societal outcomes.}

\added{Our analysis shows that polarization can emerge without algorithmic bias or selective amplification. Even when platforms treat users symmetrically and search frictions are uniform, endogenous decisions about content production and following can generate highly polarized information environments. This challenges the conventional view that online polarization stems primarily from algorithmic manipulation or filter bubbles. As a result, interventions that merely adjust exposure—such as debiasing recommendations or promoting diverse content in feeds—are likely ineffective. Effective policies must instead target the incentives for content creation.}

\added{Our framework abstracts from several features. We assume complete knowledge of costs and benefits; introducing incomplete information would allow studying how uncertainty affects network formation and welfare. Our model is static; incorporating preference evolution explicitly would formalize how today's network structure shapes tomorrow's polarization. Finally, we focus on a single platform; studying competition between platforms with different design choices and monetization strategies is a promising direction for future work.}

\newpage

\setcounter{theorem}{0}
\setcounter{proposition}{0}
\setcounter{corollary}{0}

\renewcommand{\theequation}{A-\arabic{equation}}
\renewcommand\theproposition{A-\arabic{proposition}}
\renewcommand\thecorollary{A-\arabic{corollary}}
\renewcommand\thetheorem{A-\arabic{theorem}}

\renewcommand{\thesubsection}{\Alph{subsection}}

\appendixtitleon
\appendixtitletocon
\begin{appendices}

\expandafter\def\expandafter\normalsize\expandafter{%
    \normalsize%
    \setlength\abovedisplayskip{-8pt}%
    \setlength\belowdisplayskip{-8pt}%
    \setlength\abovedisplayshortskip{-8pt}%
    \setlength\belowdisplayshortskip{-8pt}%
}
\section{Proofs}\label{appendixA}

\begin{proof}[Proof of Lemma \ref{lemma:BK}] To prove part \textit{(i)}, suppose \textit{ad absurdum} $x^{\ast}_{i}+\bar{x}^{\ast}_{i}(\mathbf{g}^\ast)<\hat{x}_{i}$. 
The assumptions on $f(\cdot)$ imply $t_{i}f'(x^{\ast}_{i}+\bar{x}^{\ast}_{i}(\mathbf{g}^\ast))>c$. 
Thus, there exists some $\epsilon >0$, such that $x^{\ast}_{i}+\epsilon+\bar{x}^{\ast}_{i}(\mathbf{g}^\ast)\leq\hat{x}_{i}$.
Then, $t_{i}[f(x^{\ast}_{i}+\epsilon+\bar{x}^{\ast}_{i}(\mathbf{g}^\ast))-f(x^{\ast}_{i}+\bar{x}^{\ast}_{i}(\mathbf{g}^\ast))]>c\epsilon$ and contributing more is profitable. Hence, $x^{\ast}_{i}$ is not an equilibrium strategy. If instead $x^{\ast}_{i}+\bar{x}^{\ast}_{i}(\mathbf{g}^\ast)>\hat{x}_{i}$ and $x^{\ast}_{i}>0$, $t_{i}f'(x^{\ast}_{i}+\bar{x}^{\ast}_{i}(\mathbf{g}^\ast))<c$. 
There exists some $\epsilon>0$ so that $x^{\ast}_{i}-\epsilon+\bar{x}^{\ast}_{i}(\mathbf{g}^\ast)\geq\hat{x}_{i}$. 
Then, $t_{i}[f(x^{\ast}_{i}+\bar{x}_{i}(\mathbf{g}^\ast))-f(x^{\ast}_{i}-\epsilon+\bar{x}^{\ast}_{i}(\mathbf{g}^\ast))]<c\epsilon$ and contributing less to $x$ is profitable. Similar arguments apply for $y$. Lemma \ref{lemma:BK} follows.
\end{proof}

\begin{proof}[Proof of Lemma \ref{lemma:zeal}]
\added{Recall: for $t_i=1$ the optimal isolation bundle is $(\hat{x}_i,0)$, for $t_i=0$ it is $(0,\hat{y}_i)$ and for all $t_i\in(0,1)$ it is $(\hat{x}_i,\hat{y}_i)=(f'^{-1}(c/t_i),f'^{-1}(c/(1-t_i)))$.}
\\
\added{\textbf{Part (i):} For $t_i\in(0,1)$, by the inverse function theorem,}
$$\added{\frac{\partial \hat{x}_i}{\partial t_i} = \frac{1}{f''(\hat{x}_i)} \cdot \frac{\partial(c/t_i)}{\partial t_i} = -\frac{c}{t_i^2 f''(\hat{x}_i)}.}$$
\added{Since $f''(\cdot) < 0$, we have $\partial \hat{x}_i/\partial t_i > 0$, so $\hat{x}_i$ is strictly increasing in $t_i$. Similarly,}
$$\added{\frac{\partial \hat{y}_i}{\partial t_i} = \frac{1}{f''(\hat{y}_i)} \cdot \frac{\partial(c/(1-t_i))}{\partial t_i} = \frac{c}{(1-t_i)^2 f''(\hat{y}_i)} < 0,}$$
\added{so $\hat{y}_i$ is strictly decreasing in $t_i$. By continuity, monotonicity extends to all $t_i\in[0,1]$.}
\\
\added{\textbf{Part (ii):} For notational convenience, write $\hat{x}(t)$ for $\hat{x}_i$ when $t_i=t$, and similarly for $\hat{y}(t)$. For $t\in(0,1)$, define total isolation demand 
$$D(t) := \hat{x}(t) + \hat{y}(t).$$
By symmetry of \eqref{payoffs} around $t = 1/2$, $D(t) = D(1-t)$ for all $t$. Differentiating,}
$$
\added{\frac{dD(t)}{dt} 
= -\frac{c}{t^2 f''(\hat{x}(t))} 
+ \frac{c}{(1-t)^2 f''(\hat{y}(t))}.}
$$
\added{Using the product rule on $\frac{dD(t)}{dt}$ and applying the chain rule with $\frac{d}{dt}\!\left(\frac{1}{f''(\hat{x}(t))}\right) = -\frac{f'''(\hat{x}(t))}{(f''(\hat{x}(t)))^2}\cdot \frac{d\hat{x}(t)}{dt}$, then substituting $\frac{d\hat{x}(t)}{dt}$ (and analogously for $\frac{d\hat{y}(t)}{dt}$), we obtain}
$$
\added{\frac{d^2D(t)}{dt^2} = -\frac{f'''(\hat{x}(t))}{(f''(\hat{x}(t)))^3} \cdot \frac{2c^2}{t^4} - \frac{f'''(\hat{y}(t))}{(f''(\hat{y}(t)))^3} \cdot \frac{2c^2}{(1-t)^4}.}
$$
\added{Since $f'' < 0$, $(f'')^3 < 0$. Then, $-\frac{f'''}{(f'')^3} = \frac{f'''}{|f''|^3}$, which has the same sign as $f'''$. Under Assumption~\ref{ass:zeal}, $f'''$ has constant sign on $[0,\infty)$, so $d^2D(t)/dt^2$ has the same sign as $f'''$ for all $t \in (0,1)$. Hence:}
\begin{enumerate}
    \item \added{If $f''' > 0$, then $D(t)$ is convex. Combined with symmetry around $t = 1/2$, this implies $D(t)$ is strictly convex in $|t - 1/2|$, reaching its minimum at $t = 1/2$. Thus, $D(0) = D(1) > D(1/2)$, which means $z(f,c) > 0$ by definition.}
    \item \added{If $f''' < 0$, then $D(t)$ is concave in $|t - 1/2|$, reaching its maximum at $t = 1/2$, so $D(0) = D(1) < D(1/2)$ and $z(f,c) < 0$.}
    \item \added{If instead $f''' = 0$, then $D(t)$ is affine and $D(0)=D(1)=D(1/2)$, so $z(f,c)=0$.}
\end{enumerate}
\added{This establishes part (ii).}
\end{proof}

\noindent For the remaining proofs, we need some definitions. There is a \textbf{path} between $i\text{ and }j$ of length $1$ if $\max\{g_{ij},g_{ji}\}=1$, and of length $m+1$ if there are $m$ players $j_{1},...,j_{m}$ distinct from $i$ and $j$, with $\max\{g_{ij_{1}},g_{j_{1}i}\}=\max\{g_{j_{1}j_{2}},g_{j_{1}j_{2}}\}=\cdot\cdot\cdot=\max\{g_{j_{m}j},g_{jj_{m}}\}=1$. We say that a set of players is a \textbf{component} of ${\bf g}$ if there is a path between every pair of them and no path to players outside the set.
\\
Given a graph ${\bf g}$, with $g_{ij}=0$, $i$ establishes a link to $j$, if $t_{i}f(x_{i}'+\bar{x}_{i}({\bf g}+ij))+(1-t_{i})f(y_{i}'+\bar{y}_{i}({\bf g}+ij))-c(x_{i}'+y_{i}')-\eta_{i}({\bf g}+ij)k\geq t_{i}f(x_{i}+\bar{x}_{i}({\bf g}))+(1-t_{i})f(y_{i}+\bar{y}_{i}({\bf g}))-c(x_{i}+y_{i})-\eta_{i}({\bf g})k$ for $(x_{i}',y_{i}')=\arg\max_{x,y\in X\times Y}U_{i}((x,x_{-i}),(y,y_{-i}),{\bf g}+ij)$ and where ${\bf g}+ij$ is the same as ${\bf g}$ setting $g_{ij}=1$. We denote the \textbf{gains of $i$ from the link to $j$} (holding other players' strategies fixed) abstracting from the cost of the link as
\begin{eqnarray*}
GL_{i}({\bf g},{\bf g}+ij)&=&c(x_{i}-x_{i}'+y_{i}-y_{i}')+t_{i}[f(x'_i+\bar{x}({\bf g}+ij))-f(x_i+\bar{x}({\bf g}))]\\&+&(1-t_{i})[f(y'_i+\bar{y}({\bf g}+ij))-f(y_i+\bar{y}({\bf g}))].
\end{eqnarray*}
Given $g_{ij}=1$ in ${\bf g}$, $i$ deletes the link to $j$, if $t_{i}f(x_{i}+\bar{x}_{i}({\bf g}))+(1-t_{i})f(y_{i}+\bar{y}_{i}({\bf g}))-c(x_{i}+y_{i})-\eta_{i}({\bf g})k< t_{i}f(x_{i}'+\bar{x}_{i}({\bf g}-ij))+(1-t_{i})f(y_{i}'+\bar{y}_{i}({\bf g}-ij))-c(x_{i}'+y_{i}')-\eta_{i}({\bf g}-ij)k$ for $(x_{i}',y_{i}')=\arg\max_{x,y\in X\times Y}U_{i}((x,x_{-i}),(y,y_{-i}),{\bf g}-ij)$ and where ${\bf g}-ij$ is as ${\bf g}$ but with $g_{ij}=0$. We denote the \textbf{loss of $i$ from deleting the link to $j$} (holding other players’ strategies fixed) abstracting from the savings in linking costs as
\begin{eqnarray*}
LL_{i}({\bf g},{\bf g}-ij)&=&c(x_{i}'-x_{i}+y_{i}'-y_{i})+t_{i}[f(x_i+\bar{x}({\bf g}))-f(x'_i+\bar{x}({\bf g}-ij))]\\&+&(1-t_{i})[f(y_i+\bar{y}({\bf g}))-f(y'_i+\bar{y}({\bf g}-ij))].
\end{eqnarray*}

\begin{proof}[Proof of Proposition \ref{prop:charact}] The proof results from a series of lemmata. \added{The first establishes equilibrium existence via a constructive argument. As for the equilibrium characterization, first we prove that no network contributor can provide less of both public goods than another network contributor (Lemma \ref{lemma3}). We then rule out structures where more than two network contributors link among themselves as equilibrium candidates (Lemma \ref{lemma4}). This allows us to rule out isolated players in a collaborative equilibrium (Lemma \ref{lemma5}) and players who free ride on multiple network contributors when there are more than two network contributors (Lemma \ref{lemma6}).}
\begin{lemma}\label{lemma:existence}
    A Nash equilibrium of the game exists.
\end{lemma}
\begin{proof}[Proof of Lemma \ref{lemma:existence}] Let ${\bf t}$ be the realization of types. If the network is empty, $(x_i,y_i)=(\hat{x}_i,\hat{y}_i)$ for all $i\in N$. Define $K$ as the gains from the most profitable link: formally, $K=GL_{i}({\bf g},{\bf g}+ij)$, where $GL_{i}({\bf g},{\bf g}+ij)\geq GL_{i'}({\bf g},{\bf g}+i'j') \text{ for all } i,j \text{ and }i',j'\in N$. If $k>K$, the empty network is an equilibrium.

\noindent Suppose $k\leq K$. To prove equilibrium existence in this case, we propose an algorithm that fixes players $1$ and $0$ as network contributors in an initial step. Moreover, we let every player sponsor a link to either $1$ or $0$ for whom doing so is profitable. Any player who does not want to free ride on $1$ or $0$ provides too little of one public good, so neither $0$ nor $1$ wants to link. Among the remaining players, if any, we can construct a similar argument by fixing the contributions of the player with the highest total isolation demand at her isolation bundle, and letting everyone establish links or reconsider their linking choice. This procedure must terminate in a bounded type space, and no profitable deviations for network contributors emerge in the process. 

\noindent Consider the following algorithm.

\textbf{Step 0.} Start with the empty network and obtain $\mathbf{g}_0$ by letting 
any player who finds it profitable to link to $1$ and $0$, i.e., set $g_{p1}=1$ ($g_{p0}=1$) if $GL_{p}({\bf g},{\bf g}+p1)=c\hat{x}_p+t_p[f(\hat{x}_1)-f(\hat{x}_p)]\geq k$ ($GL_{p}({\bf g},{\bf g}+p0)=c\hat{y}_p+(1-t_p)[f(\hat{y}_0)-f(\hat{y}_p)]\geq k)$. By equation \eqref{payoffs}, we have: \textit{(i)}, in any equilibrium $g^\ast_{01}=g^\ast_{10}=0$ as $GL_0(\mathbf{g},\mathbf{g}+01)=GL_1(\mathbf{g},\mathbf{g}+10)=0<k$; \textit{(ii)}, if $GL_{i}({\bf g},{\bf g}+i1)=c\hat{x}_i+t_i[f(\hat{x}_1)-f(\hat{x}_i)]<k$ for some player $i$, then $GL_{1}({\bf g},{\bf g}+1i)=c\hat{x}_i<k$; a similar argument applies for $0$. So, there is no profitable deviation for players $0$ and $1$. As no player contributes more to the public good than them (Lemma \ref{lemma:zeal}.(i)), and only those not linking to both are active, if all others sponsor a link, this is an equilibrium. Otherwise, go to Step 1.

\textbf{Step 1.} 
Take $i\in N\setminus \{C(\mathbf{g}_0)\cup P(\mathbf{g}_0)\}$ such that for all $i,j\in N\setminus \{C(\mathbf{g}_0)\cup P(\mathbf{g}_0)\}$, $\hat{x}_i+\hat{y}_i\geq\hat{x}_j+\hat{y}_j$. If for some players $i,j\in N\setminus \{C(\mathbf{g}_0)\cup P(\mathbf{g}_0)\}$, $\hat{x}_i+\hat{y}_i=\hat{x}_j+\hat{y}_j$, take the player with more extreme taste, i.e., $\vert t_i-1/2\vert>\vert t_j-1/2\vert$ (or one at random if $=$). Fix $i$'s contributions at $(x_i,y_i)=(\hat{x}_i,\hat{y}_i)$ and define $C(\mathbf{g}_1)=C(\mathbf{g}_0)\cup\{i\}$. 
\\
Let each $j\in N\setminus C(\mathbf{g}_1)$ establish their profitable links. 
By construction, $P(\mathbf{g}_1)\supseteq P(\mathbf{g}_0)$ as any $j\in P(\mathbf{g}_0)$ can link to the same player. If $N\setminus\{C(\mathbf{g}_1)\cup P(\mathbf{g}_1)\}=\emptyset$, there are no isolated players and the proposed configuration is an equilibrium. Otherwise, repeat step 1 until there are no isolated players, and an equilibrium is reached.

The method for identifying network contributors at each iteration of Step 1 ensures that no profitable deviation emerges for network contributors as, for any $h\in C(\mathbf{g}_m)$, $j\not\in C(\mathbf{g}_m)$ and $j\in C(\mathbf{g}_{m+1})$, $GL_j(\mathbf{g}_m,\mathbf{g}_m+jh)=c(\min\{\hat{x}_j,\hat{x}_h\}+\min\{\hat{y}_j,\hat{y}_h\})+t_j\max\{[f(\hat{x}_h)-f(\hat{x}_j)],0\}+(1-t_j)\max\{[f(\hat{y}_h)-f(\hat{y}_j)],0\}\geq GL_h(\mathbf{g}_m,\mathbf{g}_m+hj)=c(\min\{\hat{x}_j,\hat{x}_h\}+\min\{\hat{y}_j,\hat{y}_h\})+t_h\max\{[f(\hat{x}_j)-f(\hat{x}_h)],0\}+(1-t_h)\max\{[f(\hat{y}_j)-f(\hat{y}_h)],0\}=GL_h(\mathbf{g}_{m+1},\mathbf{g}_{m+1}+hj)$. This concludes the proof.
\hfill\end{proof}

\noindent Next, we establish that among network contributors, higher provision on one dimension must be offset by lower provision on the other.

\begin{lemma}\label{lemma3}
For any $i,j\in C(\mathbf{g}^\ast)$, \added{if and only if $x^{\ast}_{i}>x^{\ast}_{j}$, then $y^{\ast}_{i}<y^{\ast}_{j}$.}
\end{lemma}
\begin{proof} Suppose wlog $i>j$ and $i,j\in C(\mathbf{g})$. Moreover, suppose \textit{ad absurdum} that $v_i^{\ast}>v_j^{\ast}$, with $v=\{x,y\}$. By assumption, $c(x^{\ast}_i+y^{\ast}_i)>c(x^{\ast}_{j}+y^{\ast}_{j})$. \added{It must hold that $c(x^{\ast}_{j}+y^{\ast}_{j}) \geq LL_p(\mathbf{g}^\ast,\mathbf{g}^\ast-pj)\geq k$ for some $p\in N$, because player $p$ accesses bundle $(x^\ast_j,y^\ast_j)$ when $g^\ast_{pj}=1$ and $j\in C(\mathbf{g}^\ast)$ by assumption}. \added{It follows that}
\begin{eqnarray*}
\added{LL_p(\mathbf{g}^\ast,\mathbf{g}^\ast-pj)=c(\min\{x^\ast_j,x'_p\}+\min\{y^\ast_j,y'_p\})+t_p[\max\{f(\bar{x}_p^\ast(\mathbf{g}^\ast))-f(\hat{x}_p),0\}]}\\
\added{+(1-t_p)[\max\{f(\bar{y}_p^\ast(\mathbf{g}^\ast))-f(\hat{y}_p),0\}]\leq c(x^\ast_j+y^\ast_j).}
\end{eqnarray*}
\added{Hence, $LL_j(\mathbf{g}^\ast,\mathbf{g}^\ast-ji)=c(x^\ast_i+y^\ast_i)>c(x^\ast_j+y^\ast_j)\geq LL_p(\mathbf{g}^\ast,\mathbf{g}^\ast-pj)$, which implies} $g^\ast_{ij}=g^\ast_{ji}=1$.
\\
Since $\hat{y}_i<\hat{y}_j$ and $y^\ast_i>0$, there must exist at least one player $h$, such that $g^\ast_{ih}=0$, $g^\ast_{jh}=1$ and $y^\ast_h>0$. Otherwise, Lemma \ref{lemma:BK} implies $\hat{y}_i=y^\ast_i+y^\ast_i(\mathbf{g}^\ast)\geq y^\ast_j+y^\ast_j(\mathbf{g}^\ast)\geq\hat{y}_j$, a contradiction. \added{We prove that no such player $h$ can exist. First, note that} since $k>0$ and $j\in C(\mathbf{g}^\ast)$, $j$ is active in at least one good. 
We distinguish three cases.

\textbf{Case 1:} $x^\ast_j, y^\ast_j>0$. Then, $h\not\in N_{i}(\mathbf{g}^\ast)$ and $h\in N_j(\mathbf{g}^\ast)$ implies $LL_{j}(\mathbf{g}^\ast,\mathbf{g}^\ast-jh)=c(x^{\ast}_{h}+y^{\ast}_{h})\geq k>c(x^{\ast}_{h}+y^{\ast}_{h})=GL_{i}(\mathbf{g}^\ast,\mathbf{g}^\ast+ih)$, a contradiction.

\textbf{Case 2:} $x^\ast_j>0$ and $y^\ast_j=0$. Clearly, $x^\ast_j>x^\ast_h$, since $GL_i(\mathbf{g}^\ast,\mathbf{g}^\ast+ih)\geq c\min\{x^\ast_i,x^\ast_h\}>cx^\ast_j=LL_i(\mathbf{g}^\ast,\mathbf{g}^\ast-ij)>k$ either implies $j\not\in C(\mathbf{g}^\ast)$, or $h\in N_i(\mathbf{g}^\ast)$, a contradiction. \added{We then distinguish three sub-cases.}

\added{\textbf{Case 2a:} $y^\ast_i\geq y^\ast_h$. This would imply} $LL_{j}(\mathbf{g}^\ast,\mathbf{g}^\ast-jh)=c(x^{\ast}_{h}+y'^{\ast}_{j})+(1-t_{j})[f(\bar{y}^{\ast}_{j}(\mathbf{g}^\ast))-f(\hat{y}_{j})]<c(x^{\ast}_{h}+y^{\ast}_{h})=GL_{i}(\mathbf{g}^\ast,\mathbf{g}^\ast+ih)<k$, since, by strict concavity of $f(\cdot)$, $cy^\ast_h>cy'^\ast_j+(1-t_j)[f(\bar{y}^\ast_i(\mathbf{g^\ast}))-f(\hat{y}_j)]$. This implies $h\not\in N_j(\mathbf{g}^\ast)$, a contradiction. 

\added{\textbf{Case 2b:} $y'^\ast_j<y^\ast_i<y^\ast_h$. Then,} it must hold that $GL_i(\mathbf{g}^\ast,\mathbf{g}^\ast+ih)=c(x^\ast_h+y^\ast_i)+(1-t_i)[f(\bar{y}^\ast_i(\mathbf{g^\ast}+ih))-f(\hat{y}_i)]<k\leq c(x^\ast_h+y'^\ast_j)+(1-t_j)[f(\bar{y}^\ast_j(\mathbf{g^\ast}))-f(\hat{y}_j)]=LL_j(\mathbf{g}^\ast,\mathbf{g}^\ast-jh)$. This is a contradiction because, \textit{(i)}, $f(\cdot)$ is strictly concave, \textit{(ii)}, \added{as $y'^{\ast}_j=\max\{y_j^\ast-y^\ast_h,0\}$}, $c(x^\ast_h+y'^\ast_j)<c(x^\ast_h+y^\ast_j)$, and, \textit{(iii)}, $\hat{y}_i$ and $\hat{y}_j$ are strictly increasing in $(1-t_i)$ and $(1-t_j)$ respectively. 

\added{\textbf{Case 2c:}} $y'^\ast_j>y^\ast_i$. Note, $x^\ast_h>0$, as otherwise $\hat{x}_i=x^\ast_i+\bar{x}_i(\mathbf{g}^\ast)=x^\ast_j+\bar{x}_j(\mathbf{g}^\ast)=\hat{x}_j$, a contradiction to Lemma \ref{lemma:BK} given that $t_i\ne t_j$. Since $i$ and $h$ are active in both goods, for each $\ell$, with $\ell\in N_h(\mathbf{g}^\ast)$, $\ell\in N_i(\mathbf{g}^\ast)$. This player exists, because $g^\ast_{ij}=1$, which also implies $g^\ast_{hj}=1$. Then, however, choosing $g_{ij}=0$ ($g_{hj}=0$) and $g_{ih}=1$ ($g_{hi}=1$) is a profitable deviation for $i$ ($h$), a contradiction \added{that rules out \textbf{Case 2}}.

\textbf{Case 3:} $x^\ast_j=0$ and $y^\ast_j>0$. Clearly, $y^\ast_j>y^\ast_h$, since $GL_i(\mathbf{g}^\ast,\mathbf{g}^\ast+ih)\geq c\min\{y^\ast_i,y^\ast_h\}>cy^\ast_j=LL_i(\mathbf{g}^\ast,\mathbf{g}^\ast-ij)\geq k$ either implies $j\not\in C(\mathbf{g}^\ast)$ or $h\in N_i(\mathbf{g}^\ast)$, a contradiction. \added{We distinguish between three sub-cases}. 

\added{\textbf{Case 3a}:} $x^\ast_i\geq x^\ast_h$. We directly reach a contradiction, since $GL_i(\mathbf{g}^\ast,\mathbf{g}^\ast+ih)=c(x^\ast_h+y^\ast_h)>c(x'^\ast_j+y_h)+t_j[f(\bar{x}^\ast_j(\mathbf{g^\ast}))-f(\hat{x}_j)]=LL_j(\mathbf{g}^\ast,\mathbf{g}^\ast-jh)$ follows from strict concavity of $f(\cdot)$. 

\added{\textbf{Case 3b}:} $x'^\ast_j<x^\ast_i<x^\ast_h$: The fact that $i>j$ directly implies $GL_i(\mathbf{g}^\ast,\mathbf{g}^\ast+ih)=c(x^\ast_i+y^{\ast}_{h})+t_{i}[f(\bar{x}^{\ast}_{i}(\mathbf{g}^\ast+ih))-f(\hat{x}_{i})]>c(x'^\ast_j+y^{\ast}_{h})+t_{j}[f(\bar{x}^{\ast}_{j}(\mathbf{g}^\ast))-f(\hat{x}_{j})]=LL_j(\mathbf{g}^\ast,\mathbf{g}^\ast-jh)$, a contradiction. 

\textbf{Case 3c}: $x'^\ast_j>x^\ast_i$: By construction, $y^\ast_h>0$. It must hold that $cy^\ast_h<k$, since $g^\ast_{ih}=1$ is a profitable deviation otherwise. This implies $x^\ast_h>0$, since $g^\ast_{jh}=0$ otherwise. Both $i$ and $h$ provide both types of information and thus, for each player $\ell$ with $\ell\in N_h(\mathbf{g}^\ast)$, $\ell\in N_i(\mathbf{g}^\ast)$. This player must exist, since $g^\ast_{ij}=1$, which implies $g^\ast_{hj}=1$. 
Then, choosing $g_{ij}=0$ ($g_{hj}=0$) and $g_{ih}=1$ ($g_{hi}=1$) is a profitable deviation for $i$ ($h$), a contradiction.

\added{Hence, there is no player $h$ such that $g^\ast_{ih}=0$, $g^\ast_{jh}=1$ and $y^\ast_h>0$,} concluding the proof of Lemma \ref{lemma3}.
\end{proof}

\noindent We now establish conditions for players in $C(\mathbf{g}^\ast)$ to link to each other in equilibrium. 
\begin{lemma}\label{lemma4}
In any Nash equilibrium, if for $i$ and $j\in C(\mathbf{g}^\ast)$, $g^{\ast}_{ij}=1$, or $g^{\ast}_{ji}=1$, or both, $i$ and $j$ are the only elements in $C(\mathbf{g}^\ast)$.
\end{lemma}
\begin{proof}
\added{Suppose without loss of generality that $i>j$ and $g^{\ast}_{ij}=1$. We show that no third network contributor can exist. To prove this, we assume by contradiction that there exists $h \in C(\mathbf{g}^\ast)$ with $h \neq i,j$. Then, we distinguish two cases: we first show that if $j$ sponsors no links, then $j$ must be $i$'s only neighbor; we then show that, given this structure, no third contributor $h$ can receive a link independently of its type.}

\medskip

\added{\textbf{Case 1: $j$ sponsors no links.} 
As $N_j(\mathbf{g}^\ast)=\emptyset$, $(x_j^\ast,y_j^\ast)=(\hat{x}_j,\hat{y}_j)$. Since $g^\ast_{ij}=1$, $y_i^\ast=0$. Since $i \in C(\mathbf{g}^\ast)$, Lemma~\ref{lemma3} implies $cx_i^\ast \geq k$ and $x_i^\ast > x_\ell^\ast$ for all $\ell \in N \setminus \{i\}$.}

\medskip

\added{\emph{\textbf{Step 1:} We establish that $j$ is $i$'s only neighbor.}}
\\
\added{Suppose by contradiction that $i$ has another neighbor $h' \neq j$, so $g^\ast_{ih'}=1$. Then }
\[
LL_i(\mathbf{g}^\ast,\mathbf{g}^\ast - ih') = cx_{h'}^\ast + (1-t_i)[f(\hat{y}_j + y_{h'}^\ast) - f(\hat{y}_j)] \geq k.
\]
\added{If $x_{h'}^\ast < \hat{x}_j$, then by Lemma~\ref{lemma3}, $y_{h'}^\ast > \hat{y}_j$. But then}
\begin{align*}
GL_j(\mathbf{g}^\ast, \mathbf{g}^\ast + jh') &= c(x_{h'}^\ast + \hat{y}_j) + (1-t_j)[f(y_{h'}^\ast) - f(\hat{y}_j)]\\
&> cx_{h'}^\ast + (1-t_i)[f(\hat{y}_j + y_{h'}^\ast) - f(\hat{y}_j)]\\
&= LL_i(\mathbf{g}^\ast, \mathbf{g}^\ast - ih') \geq k,
\end{align*}
\added{where the inequality follows from $t_i > t_j$ and the concavity of $f$. This implies $g^\ast_{jh'}=1$, contradicting $N_j(\mathbf{g}^\ast) = \emptyset$. Hence, $x_{h'}^\ast > \hat{x}_j$ and $y_{h'}^\ast < \hat{y}_j$ for any neighbor $h'$ of $i$ other than $j$. We now show this is impossible for any type $h'\in N$:}
\begin{itemize}
\item \added{If $h' > i$: Player $h'$ must link to some player $\ell \neq j$ to whom $i$ does not link (otherwise $x_i^\ast < x_{h'}^\ast$, contradicting Lemma~\ref{lemma3}). Since $h' > i > j$, we have $\ell > j$ (else $g^\ast_{h'\ell}=0$ by revealed preference). Repeating this argument generates a strictly descending chain of types. Since the type space is bounded, the chain must terminate at player 1, where Lemma \ref{lemma3} is violated.}
\item \added{If $j < h' < i$: Then}
\[
GL_{h'}(\mathbf{g}^\ast, \mathbf{g}^\ast + h'j) = c(\hat{x}_j + \hat{y}_{h'}) + (1-t_{h'})[f(\hat{y}_j) - f(\hat{y}_{h'})] > LL_i(\mathbf{g}^\ast, \mathbf{g}^\ast - ij) \geq k,
\]
\added{where the inequality follows from $t_i > t_{h'}$ and $\hat{y}_i < \hat{y}_{h'}$. This implies $g^\ast_{h'j}=1$, so $y_{h'}^\ast = 0$. But then $y_{h'}^\ast = y_i^\ast = 0$, contradicting Lemma~\ref{lemma3}.}
\item \added{If $h' < j$: Then $\hat{x}_{h'} < \hat{x}_j$ by Lemma~\ref{lemma:zeal}(i), so $x_{h'}^\ast < \hat{x}_j$, contradicting $x_{h'}^\ast > \hat{x}_j$.}
\end{itemize}
\added{Hence $j$ is $i$'s only neighbor, i.e., $N_i(\mathbf{g}^\ast) = \{j\}$. By a similar argument, any other contributor $h$ can have at most one neighbor.}

\medskip

\added{\emph{\textbf{Step 2:} We show that $h \in C(\mathbf{g}^\ast)$ leads to a contradiction for any location of $h$.}}

\added{\textbf{Step 2-Case a: $h > i$.}
Then $g^\ast_{hi}=1$ since $GL_h(\mathbf{g}^\ast, \mathbf{g}^\ast + hi) = cx_i^\ast \geq k$ and $x_i^\ast \geq x_\ell^\ast$ for all $\ell \in N$. 
By Lemma~\ref{lemma3}, $j < i < h$ implies $x_i^\ast > x_h^\ast > x_j^\ast$ and $0 < y_h^\ast < y_j^\ast$. Since $h \in C(\mathbf{g}^\ast)$, we have $c(x_h^\ast + y_h^\ast) \geq k$. 
We claim $g^\ast_{hj}=0$. If $g^\ast_{hj}=1$, then $y_h^\ast = 0$, contradicting Lemma~\ref{lemma3} since $y_i^\ast = 0$ as well.}

\added{Now consider player $1$. Since $GL_1(\mathbf{g}^\ast, \mathbf{g}^\ast + 1i) = cx_i^\ast \geq k$, we have $g^\ast_{1i}=1$. Similarly, $g^\ast_{1h}=1$ is profitable if $cx_h^\ast \geq k$. But $g^\ast_{ih}=0$ (since $N_i(\mathbf{g}^\ast) = \{j\}$), so by Lemma~\ref{lemma:zeal}(i), player $1$ does not link to $h$.}

\added{For any $q \in (0,i)$, since $g^\ast_{ij}=1$, $g^\ast_{qj}=1$ by Lemma~\ref{lemma:zeal}(i). And since $g^\ast_{ih}=0$, we have $g^\ast_{qh}=0$. 
Therefore, no player $p \in N$ satisfies $g^\ast_{ph}=1$, contradicting $h \in C(\mathbf{g}^\ast)$.}

\added{\textbf{Step 2-Case b: $j < h < i$.}
Then}
\[
GL_h(\mathbf{g}^\ast, \mathbf{g}^\ast + hj) = c(\hat{x}_j + \hat{y}_h) + (1-t_h)[f(\hat{y}_j) - f(\hat{y}_h)] > LL_i(\mathbf{g}^\ast, \mathbf{g}^\ast - ij) \geq k,
\]
\added{where the inequality follows from $t_i > t_h$ and $\hat{y}_i < \hat{y}_h$. This implies $g^\ast_{hj}=1$, so $y_h^\ast \leq y_i^\ast = 0$ and $x_h^\ast < x_i^\ast$ by Lemma~\ref{lemma3}. But then $h$ does not receive links, contradicting $h \in C(\mathbf{g}^\ast)$.}

\added{\textbf{Step 2-Case c: $h < j$.}
We show that player $0$ cannot receive links. For any $h \in (0,j)$,}
\[
GL_0(\mathbf{g}^\ast, \mathbf{g}^\ast + 0j) = c\hat{y}_j < c(\hat{x}_h + \hat{y}_j) + t_h[f(\hat{x}_j) - f(\hat{x}_h)],
\]
\added{so linking to any $h \in (0,j)$ dominates linking to $0$. Hence $0$ receives no links.}

\added{Since $0$ can sponsor at most one link, it suffices to show $g^\ast_{0j}=0$. But if $g^\ast_{0j}=1$, then since $j$ sponsors no links,}
\[
GL_j(\mathbf{g}^\ast, \mathbf{g}^\ast + j0) = c(\hat{y}_0 - \hat{y}_j) < k \leq c\hat{y}_j,
\]
\added{where the first inequality holds as otherwise $j$ would sponsor links. This contradicts $0 \in C(\mathbf{g}^\ast)$. 
By similar arguments for any $h < j$, no such $h$ can be in $C(\mathbf{g}^\ast)$.}

\medskip

\added{\textbf{Case 2: $j$ sponsors some links.} 
We show that network contributors sponsor the same number of links, and then rule out the case where they sponsor any links.}

\added{\textbf{Case 2a: Suppose some contributors sponsor different numbers of links.} 
Suppose without loss of generality that $|N_i(\mathbf{g}^\ast)| > |N_j(\mathbf{g}^\ast)|$. 
By Lemma~\ref{lemma3}, at most two players in $C(\mathbf{g}^\ast)$ can be inactive in one good. For every $\ell \in N_i(\mathbf{g}^\ast) \setminus N_j(\mathbf{g}^\ast)$, either $x_\ell^\ast > x_j^\ast$ or $y_\ell^\ast > y_j^\ast$. 
Since $i > j$ and $i$ sponsors more links, we have}
\[
x_i^\ast = \hat{x}_i - \hat{x}_j + \sum_{\ell \in N_j(\mathbf{g}^\ast)} x_\ell^\ast - \sum_{\ell \in N_i(\mathbf{g}^\ast) \setminus \{j\}} x_\ell^\ast.
\]
\added{If $y_i^\ast > 0$, then $\sum_{\ell \in N_j(\mathbf{g}^\ast)} y_\ell^\ast > \sum_{\ell \in N_i(\mathbf{g}^\ast)} y_\ell^\ast$ (since $j$ sponsors fewer links and $\hat{y}_j > \hat{y}_i$). This implies $y_i^\ast < 0$, a contradiction. Hence $y_i^\ast = 0$.}
\\
\added{By Lemma \ref{lemma3}, $x_i^\ast > x_p^\ast$ for all $p \in N$. Let $i_q$ denote the neighbor of $i$ contributing the most to $x$ (among $i$'s neighbors). Then $g^\ast_{i_q i}=1$ by equilibrium.}
\\
\added{From equilibrium conditions, $\hat{x}_i - \hat{x}_{i_q} = \sum_{\ell \in N_i, \ell \neq i_q} x_\ell^\ast - \sum_{\ell \in N_{i_q}, \ell \neq i} x_\ell^\ast$. This implies $t_i > t_{i_q}$, so if $g^\ast_{im}=1$, then $g^\ast_{i_q m}=1$. Hence $N_i(\mathbf{g}^\ast) \subseteq N_{i_q}(\mathbf{g}^\ast) \cup \{i_q\}$, implying $\hat{x}_i = \hat{x}_{i_q}$, contradicting $t_i \neq t_{i_q}$.}

\added{\textbf{Case 2b: All contributors sponsor the same number of links.}
Since $i > j$ and both sponsor the same number of links, we have $\sum_{\ell \in N_j(\mathbf{g}^\ast)} y_\ell^\ast > \sum_{\ell \in N_i(\mathbf{g}^\ast)} y_\ell^\ast$ (else $i$ could profitably imitate $j$'s links). This implies $\sum_{\ell \in N_i(\mathbf{g}^\ast)} x_\ell^\ast > \sum_{\ell \in N_j(\mathbf{g}^\ast)} x_\ell^\ast$.}
\\
\added{Since $i > j$ and all players establish the same number of links, there exists a player $q$ who is inactive in $x$. For some $q' > q$ with $q' \in N_q(\mathbf{g}^\ast)$, we must have $N_q(\mathbf{g}^\ast) \setminus \{q'\} = N_{q'}(\mathbf{g}^\ast) \setminus \{q\}$ (they link to the same players). 
Then $y_q^\ast = \hat{y}_q - \hat{y}_{q'} + \sum_{m \in N_{q'}(\mathbf{g}^\ast)} y_m^\ast - \sum_{m \in N_q(\mathbf{g}^\ast)} y_m^\ast$ implies $\hat{y}_q = \hat{y}_{q'}$, contradicting $q \neq q'$.}

\added{In all cases, the existence of $h \in C(\mathbf{g}^\ast)$ with $h \neq i,j$ leads to a contradiction. Therefore, $C(\mathbf{g}^\ast) = \{i,j\}$. This concludes the proof of Lemma \ref{lemma4}.}
\end{proof}

\noindent Next, we show there are no isolated players in a collaborative equilibrium.
\begin{lemma}\label{lemma5}
If $g^{\ast}_{ij}=1$ for $i,j\in C(\mathbf{g}^\ast)$, there are no isolated players.
\end{lemma}
\begin{proof} By Lemma \ref{lemma4}, $C(\mathbf{g}^\ast)=\{i,j\}$.
Suppose wlog $g^{\ast}_{ij}=1$ and $g^{\ast}_{ji}=0$. By Lemma \ref{lemma:BK}, $(x^{\ast}_{i},y^{\ast}_{i})=(\hat{x}_{i}-\hat{x}_{j},0)$ and $(x^{\ast}_{j},y^{\ast}_{j})=(\hat{x}_{j},\hat{y}_{j})$. 

\noindent For $j<p<i$, $LL_{p}(\mathbf{g}^\ast,\mathbf{g}^\ast-pj)=c(\hat{x}_{j}+\hat{y}_{p})+(1-t_{p})[f(\hat{y}_{j})-f(\hat{y}_{p})]>c(\hat{x}_{j}+\hat{y}_{i})+(1-t_{i})[f(\hat{y}_{j})-f(\hat{y}_{i})]\geq k$ and $g^{\ast}_{pj}=1$.

\noindent Suppose $p<j$. It is sufficient to show that $0$ does not remain isolated, since $LL_{0}(\mathbf{g}^\ast,\mathbf{g}^\ast-0j)=c\hat{y}_{j}<c\hat{y}_{j}+t_{h}[f(\hat{x}_{j})-f(\hat{x}_{h})]=GL_h(\mathbf{g}^\ast,\mathbf{g}^\ast+hj)$ for all $h\in(0,j)$. The proof of Lemma \ref{lemma4} establishes that $0$ cannot be isolated when $g^\ast_{ij}=1$, since $g^\ast_{j0}=1$ would be a profitable deviation.

\noindent Suppose $p>i$. It is sufficient to show that $1$ does not remain isolated, since $LL_{1}(\mathbf{g}^\ast,\mathbf{g}^\ast-1i)=cx^{\ast}_{i}\geq LL_{h}(\mathbf{g}^\ast,\mathbf{g}^\ast-hi)$ for all $h\in N$. Since $i\in C(\mathbf{g}^\ast)$, $cx^\ast_i\geq k$, which directly implies $g^\ast_{1i}=1$. This concludes the proof of Lemma \ref{lemma5}.
\end{proof}
\begin{lemma}\label{lemma6}
If $i,j\in C(\mathbf{g}^\ast)$ and $g^{\ast}_{ij}=g^{\ast}_{ji}=0$, either $i$ and $j$ are in different components of $\mathbf{g}^\ast$, or $C(\mathbf{g}^\ast)=\{i,j\}$.
\end{lemma}
\begin{proof}
\added{Suppose there exists $h\in C(\mathbf{g}^\ast)$, where $h\neq i\neq j$.
Players $i$ and $j$ are in the same component of $\mathbf{g}^\ast$ if, and only if, there exists a player $p$ with $g^{\ast}_{pi}=g^{\ast}_{pj}=1$. Clearly, $i>p>j$.
Otherwise, either $g^{\ast}_{ij}=1\text{, or }g^{\ast}_{ji}=1$, and by Lemma \ref{lemma4} $i\text{ and }j$ would be the only elements in $C(\mathbf{g}^\ast)$.
As $g^{\ast}_{pj}=1$, $LL_{p}(\mathbf{g}^\ast,\mathbf{g}^\ast-pi)=c(\hat{x}_{p}-\hat{x}_{j})+t_{p}[f(\hat{x}_{i}+\hat{x}_{j})-f(\hat{x}_{p})]+(1-t_{p})[f(\hat{y}_{i}+\hat{y}_{j})-f(\hat{y}_{p})]$.
Similarly, as $g^{\ast}_{pi}=1$, $LL_{p}(\mathbf{g}^\ast,\mathbf{g}^\ast-pj)=c(\hat{y}_{p}-\hat{y}_{i})+t_{p}[f(\hat{x}_{i}+\hat{x}_{j})-f(\hat{x}_{p})]+(1-t_{p})[f(\hat{y}_{i}+\hat{y}_{j})-f(\hat{y}_{p})]$.
When $i>h>j$, this directly implies $g^{\ast}_{hi}=1$ or $g^{\ast}_{hj}=1$, a contradiction to Lemma \ref{lemma4}.
We therefore assume $h=1>i$ wlog.} 
Given $g^{\ast}_{1i}=0$, $g^{\ast}_{ji}=0$, and $g^{\ast}_{ij}=0$, 
\begin{eqnarray*}
    GL_{i}(\mathbf{g}^\ast,\mathbf{g}^\ast+i1)&=& c\hat{x}_{i}+t_{i}[f(\hat{x}_{1})-f(\hat{x}_{i})]<k\\
    GL_{i}(\mathbf{g}^\ast,\mathbf{g}^\ast+ij)&=& c(\hat{x}_{j}+\hat{y}_i)+(1-t_{i})[f(\hat{y}_{j})-f(\hat{y}_{i})]<k,\\
    GL_{j}(\mathbf{g}^\ast,\mathbf{g}^\ast+ji)&=& c(\hat{x}_{j}+\hat{y}_i)+t_{j}[f(\hat{x}_{i})-f(\hat{x}_{j})]<k.
\end{eqnarray*}
Furthermore, by assumption, 
\begin{eqnarray*}
    LL_{p}(\mathbf{g}^\ast,\mathbf{g}^\ast-pi)&=&c(\hat{x}_{p}-\hat{x}_{j})+t_{p}[f(\hat{x}_{i}+\hat{x}_{j})-f(\hat{x}_{p})]+(1-t_{p})[f(\hat{y}_{i}+\hat{y}_{j})-f(\hat{y}_{p})]\geq k\\
    LL_{p}(\mathbf{g}^\ast,\mathbf{g}^\ast-pj)&=&c(\hat{y}_{p}-\hat{y}_{i})+t_{p}[f(\hat{x}_{i}+\hat{x}_{j})-f(\hat{x}_{p})]+(1-t_{p})[f(\hat{y}_{i}+\hat{y}_{j})-f(\hat{y}_{p})]\geq k.
\end{eqnarray*}
Hence, 
\begin{eqnarray*}
    &&GL_{i}(\mathbf{g}^\ast,\mathbf{g}^\ast+ij)+GL_{i}(\mathbf{g}^\ast,\mathbf{g}^\ast+i1)=\\
    &=&c(\hat{x}_{j}+\hat{x}_i+\hat{y}_i)+(1-t_{i})[f(\hat{y}_{j})-f(\hat{y}_{i})]+t_{i}[f(\hat{x}_{1})-f(\hat{x}_{i})]\\
    &<&c(\hat{x}_p-\hat{x}_j+\hat{y}_p-\hat{y}_i)+t_{p}[f(\hat{x}_{i}+\hat{x}_{j})-f(\hat{x}_{p})]+(1-t_{p})[f(\hat{y}_{i}+\hat{y}_{j})-f(\hat{y}_{p})]\\
    &=&LL_{p}(\mathbf{g}^\ast,\mathbf{g}^\ast-pi)+LL_{p}(\mathbf{g}^\ast,\mathbf{g}^\ast-pj).
\end{eqnarray*}
As $f(\cdot)$ is strictly concave, $LL_{p}(\mathbf{g}^\ast,\mathbf{g}^\ast-pi)+LL_{p}(\mathbf{g}^\ast,\mathbf{g}^\ast-pj)$. By Lemma \ref{lemma:zeal}, 
$$c(\hat{x}_i+\hat{y}_i)>c(\hat{x}_p-\hat{x}_j+\hat{y}_p-\hat{y}_i)+t_p[f(\hat{x}_i+\hat{x}_j)-f(\hat{x}_p)]+(1-t_p)[f(\hat{y}_i+\hat{y}_j)-f(\hat{y}_p)],$$ because $\hat{x}_q=f'^{-1}(c/t_q)$ and $\hat{y}_q=f'^{-1}(c/(1-t_q))$ for $q\in N$ are functions of $c$. This is a contradiction which concludes the proof of Lemma \ref{lemma6}.
\end{proof}

\noindent These lemmata prove Proposition \ref{prop:charact}.
\end{proof}

\begin{proof}[Proof of Proposition \ref{prop:lotf}] By Proposition \ref{prop:charact}, in any collaborative equilibrium $\vert C(\mathbf{g}^\ast)\vert=2$ for any number of players $N$, so the statement follows in this case.
\\
For independent equilibria, there can be many network contributors. If $\vert C(\mathbf{g}^\ast)\vert< 2$, the statement trivially follows again. 
Otherwise, by Proposition \ref{prop:charact} all players $p\in P(\mathbf{g}^\ast)$ sponsor exactly one link and the equilibrium network consists of a collection of components. For any $i\in C(\mathbf{g}^\ast)$, there exists $h$, such that $$GL_h(\mathbf{g}^\ast,\mathbf{g}^\ast+hi)=c(\hat{x}_{h}+\hat{y}_i)+t_h[f(\hat{x}_i)-f(\hat{x}_{h})]\geq k,$$ where $h<i$ is wlog. Define $\bar{t}_i$ and $\underline{t}_i$ such that $$GL_p({\bf g},{\bf g}+pi)=c(\hat{x}_i+\hat{y}_p)+(1-t_p)[f(\hat{y}_i)-f(\hat{y}_p)]=k$$ if $t_p=\bar{t}_i$ and $$GL_p({\bf g},{\bf g}+pi)=c(\hat{x}_p+\hat{y}_i)+t_p[f(\hat{x}_i)-f(\hat{x}_p)]=k$$ if $t_p=\underline{t}_i$. Thus, for each $i\in C(\mathbf{g}^\ast)$, there exists $\underline{t}_{i}<t_{i}<\bar{t}_{i}$ such that for each $p$, with $t_p\in [\underline{t}_{i},\bar{t}_{i}]$, $g^\ast_{pi}=1$. Suppose \textit{ad absurdum} that $\vert C(\mathbf{g}^\ast)\vert\to\infty$. As the type space is bounded, this would imply $t_i-t_{j}\to0$, where $j<i$ wlog and $j\in C(\mathbf{g}^\ast)$. Then, $GL_{j}(\mathbf{g}^\ast,\mathbf{g}^\ast+ji)\to c(\hat{x}_{i}+\hat{y}_{i})$, so that $j\in [\underline{t}_i,\bar{t}_{i}]$ and $g^\ast_{ji}=1$, a contradiction. This concludes the proof of Proposition \ref{prop:lotf}.
\hfill \end{proof}

\begin{proof}[Proof of Proposition \ref{prop:eq_net}] In Lemma \ref{lemma:existence}, we construct an independent equilibrium for an arbitrary $k\in(0,K]$. Existence of an independent equilibrium for any $k\leq K$ follows directly. Here, we address 
collaborative equilibria. \added{To do so, we separately address collaborative equilibria where both network contributors free ride (\textbf{Case 1}) and where one network contributor provides their isolation bundle (\textbf{Case 2}).}

\noindent Suppose $i,j\in C(\mathbf{g}^\ast)$, where $i>j$ and $g^\ast_{ji}=1$ wlog. \added{In a large society, we can focus on threshold types to derive the conditions for a collaborative equilibrium to exist.} 

\noindent \textbf{Case 1:} $g^{\ast}_{ij}=1$. Proposition \ref{prop:charact} implies $(x^{\ast}_{i},y^{\ast}_{i})=(\hat{x}_{i},0)$ and $(x^{\ast}_{j},y^{\ast}_{j})=(0,\hat{y}_{j})$. A collaborative equilibrium with $i,j\in C(\mathbf{g}^\ast)$ exists if, and only if, $$LL_{i}(\mathbf{g}^\ast,\mathbf{g}^\ast-ij)=c\hat{y}_{i}+(1-t_{i})[f(\hat{y}_{j})-f(\hat{y}_{i})]\geq k>c(\hat{y}_{0}-\hat{y}_{j})=GL_{j}(\mathbf{g}^\ast,\mathbf{g}^\ast+j0),$$ \added{where $k>c(\hat{y}_{0}-\hat{y}_{j})=GL_{j}(\mathbf{g}^\ast,\mathbf{g}^\ast+j0)$ ensures $g^\ast_{j0}=0$ given that $g^\ast_{0j}=1$}. Similarly, $$LL_{j}(\mathbf{g}^\ast,\mathbf{g}^\ast-ji)=c\hat{x}_{j}+t_{j}[f(\hat{x}_{i})-f(\hat{x}_{j})]\geq k>c(\hat{x}_{1}-\hat{x}_{i})=GL_{i}(\mathbf{g}^\ast,\mathbf{g}^\ast+i1),$$ \added{where $k>c(\hat{x}_{1}-\hat{x}_{i})=GL_{i}(\mathbf{g}^\ast,\mathbf{g}^\ast+i1)$ ensures $g^\ast_{i1}=0$ given that $g^\ast_{1i}=1$}. Take $i$ and $j$ so that $i$ and $j$ are the most extreme network contributors who link to each other. In a large society, it is sufficient to consider the conditions for one agent. A collaborative equilibrium exists if $\underline{k}_{C'}<k\leq\bar{k}_{C'}$, where $\bar{k}_{C'}=c\hat{x}_{j}+t_{j}[f(\hat{x}_{i})-f(\hat{x}_{j})]$ and $\underline{k}_{C'}=c(\hat{x}_{1}-\hat{x}_{i})$ for some $i\text{ and }j\in N$. By Lemma \ref{lemma:zeal}.(ii), $\underline{k}_{C'}=c(\hat{x}_{1}-\hat{x}_{i})$ is increasing in zeal because Assumption \ref{ass:zeal} imposes monotonicity in zeal in the intervals $[0,1/2]$ and $[1/2,1]$. Moreover, the value of $\bar{k}_{C'}=c\hat{x}_{j}+t_{j}[f(\hat{x}_{i})-f(\hat{x}_{j})]$ is decreasing in zeal. Hence, there exists a threshold value $ z_{C'}$ such that no collaborative equilibrium where the network contributors link to each other exists for $z(f,c)> z_{C'}$.

\noindent \textbf{Case 2:} $g^{\ast}_{ij}=0$. Proposition \ref{prop:charact} implies $(x^{\ast}_{i},y^{\ast}_{i})=(\hat{x}_{i},\hat{y}_{i})$ and $(x^{\ast}_{j},y^{\ast}_{j})=(0,\hat{y}_{j}-\hat{y}_{i})$. A collaborative equilibrium where one core player sponsors no links exists if $$LL_{j}(\mathbf{g}^\ast,\mathbf{g}^\ast-ji)=c(\hat{x}_{j}+\hat{y}_{i})+t_{j}[f(\hat{x}_{i})-f(\hat{x}_{j})]\geq k>c(\hat{x}_1-\hat{x}_i)=GL_i(\mathbf{g}^\ast,\mathbf{g}^\ast+i1).$$ Then, $g^\ast_{0j}=1$ and $g^\ast_{j0}=0$ imply $$LL_{0}(\mathbf{g}^\ast,\mathbf{g}^\ast-0j)=c(\hat{y}_{j}-\hat{y}_{i})\geq k>c(\hat{y}_{0}-\hat{y}_{j}+\hat{y}_{i})=GL_{j}(\mathbf{g}^\ast,\mathbf{g}^\ast+j0),$$ \added{where $k>c(\hat{y}_{0}-\hat{y}_{j}+\hat{y}_{i})=GL_{j}(\mathbf{g}^\ast,\mathbf{g}^\ast+j0)$ ensures $g^\ast_{j0}=0$ given $g^\ast_{0j}=1$ at the induced equilibrium contribution of $0$, $(x^\ast_0,y^\ast_0)=(0,\hat{y}_0-\hat{y}_j+\hat{y}_i)$. Note, $\hat{y}_0-\hat{y}_j+\hat{y}_i>\hat{y}_j-\hat{y}_i$ is implied by Lemma \ref{lemma3}.} Combining conditions implies $$LL_{j}(\mathbf{g}^\ast,\mathbf{g}^\ast-ji)=c(\hat{x}_{j}+\hat{y}_{i})+t_{j}[f(\hat{x}_{i})-f(\hat{x}_{j})]\geq k>c(\hat{y}_{0}-\hat{y}_{j}+\hat{y}_{i})=GL_{j}(\mathbf{g}^\ast,\mathbf{g}^\ast+j0).$$ 
A collaborative equilibrium where one core player sponsors no links exists if, and only if, $\underline{k}_{C''}<k\leq\bar{k}_{C''}$, where $\bar{k}_{C''}=\min\{c(\hat{y}_{j}-\hat{y}_{i}),c(\hat{x}_{j}+\hat{y}_{i})+t_{j}[f(\hat{x}_{i})-f(\hat{x}_{j})]\}$ and $\underline{k}_{C''}=\max\{c(\hat{y}_0-\hat{y}_j+\hat{y}_i),c(\hat{x}_1-\hat{x}_i)\}$ for some $i\text{ and }j\in N$. A similar argument holds for $g^{\ast}_{ij}=1\text{ and }g^{\ast}_{ji}=0$. By a similar argument as in \textbf{Case 1}, if Assumption \ref{ass:zeal} holds, Lemma \ref{lemma:zeal}.(ii) implies that there exists $ z_{C''}$, such that no collaborative equilibrium where one network contributors sponsors no link exists if $z(f,c)> z_{C''}$. 

\noindent Define $\bar{k}_C=\bar{k}_{C'}$, $\underline{k}_C=\underline{k}_{C'}$, and $z_C=z_{C'}$ if the conditions in \textbf{Case 1} are less stringent than in \textbf{Case 2} and $\bar{k}_C=\bar{k}_{C''}$, $\underline{k}_C=\underline{k}_{C''}$, and $z_C=z_{C''}$ otherwise. 
\added{Hence, there must exist thresholds $\bar{k}_C$, $\underline{k}_C$, and $z_C$ such that a collaborative equilibrium exists if, and only if, $k\in[\underline{k}_{C},\bar{k}_C]$ and $z(f,c)\leq\underline{z}_C$.} This concludes the proof of Proposition \ref{prop:eq_net}.
\end{proof}

\noindent Before studying the welfare-maximizing equilibrium, we present a preliminary result.

\begin{lemma}\label{lemma:distribution}
    If $0<k<K$, Assumption \ref{ass:zeal} holds and $m\in C(\mathbf{g}^\ast)$ is the most moderate network contributor, the interval of players who link to $m$, $[\underline{t}_m,\bar{t}_m]$, narrows with zeal.
\end{lemma}

\begin{proof}
    There is $z(f,c)$ such that $\hat{x}_m+\hat{y}_m>\hat{x}_i+\hat{y}_i$ for all $i\in C(\mathbf{g}^\ast)\cup I(\mathbf{g}^\ast)$. Take some $j<m$ with $g_{jm}=1$. Then $$LL_j(\mathbf{g}^\ast,\mathbf{g}^\ast-jm)=c(\hat{x}_j+\hat{y}_m)+t_j[f(\hat{x}_m)-f(\hat{x}_j)]\geq k.$$ The LHS of the equilibrium condition decreases in zeal by Lemma \ref{lemma:zeal}.(ii) and fewer types want to link to $m$, \added{as, under Assumption \ref{ass:zeal}, zeal increases the difference between player $m$'s total isolation demand and that of other network contributors.} A similar argument holds when $j$ sponsors multiple links, so the statement follows.
\end{proof}

\begin{proof}[Proof of Proposition \ref{prop:welfare-max}] 
\added{The proof is as follows. First, Lemma \ref{lemma:welfare-max} shows that a collaborative equilibrium where both core players free ride is not welfare maximizing. Lemma \ref{lemma:PC} shows that, with intermediate zeal and linking cost, there exist types who can be network contributors who do not free ride only in a collaborative equilibrium. The welfare-maximizing equilibrium can thus be collaborative. Lemmas \ref{lemma:zeal_welfare-max} and \ref{lemma:linking-cost_welfare-max} derive necessary conditions for this to happen. Finally, Lemma \ref{lemma:xi} shows that these necessary conditions are also sufficient.}
\begin{lemma}\label{lemma:welfare-max}
    In a large society, if the welfare-maximizing equilibrium is collaborative, one core player sponsors no link. 
\end{lemma}
\begin{proof}
    Suppose \textit{ad absurdum} that $\mathbf{s}^W$ is a collaborative equilibrium where both network contributors sponsor a link, i.e., $C(\mathbf{g}^W)=\{i,j\}$ and $g^W_{ij}=g^W_{ji}=1$, where $i>j$ wlog. By Lemma \ref{lemma:zeal}.(i), the demand for a good is strictly increasing in a player's taste for it and there exists an independent equilibrium $\mathbf{s}^\ast$ such that $C(\mathbf{g}^\ast)=\{0,1\}$. For each $h$ for whom $g^W_{hi}=1$ ($g^W_{hj}=1$), $g^\ast_{h1}=1$ ($g^\ast_{h0}=1$) \added{because $\hat{x}_1>\hat{x}_i$ (and $\hat{y}_0>\hat{y}_j$). Hence, each $h\in P(\mathbf{g}^W)$ receives a higher utility in the independent equilibrium if they sponsor the same number of links, and thus incur the same cost for sponsoring links. If player $h$ sponsored more links in the independent equilibrium, it must hold that $LL_h(\mathbf{g}^\ast,\mathbf{g}^\ast-h1)\geq k$ and $LL_h(\mathbf{g}^\ast,\mathbf{g}^\ast-h0)\geq k$. Hence, welfare must indeed be higher in the independent equilibrium}. Since $\hat{x}_1>\hat{x}_i$ and $\hat{y}_0>\hat{y}_j$, $\sum_{\ell\in N} U_\ell(\mathbf{s}^W)<\sum_{\ell\in N} U_\ell(\mathbf{s}^\ast)$, a contradiction. 
    \added{By Proposition \ref{prop:lotf}, the utilities of network contributors $i$ and $j$ or $1$ and $0$ are negligible compared to that of free riders.}
\end{proof}
\noindent We now prove a lemma on zeal and the gains from linking to a moderate contributor.
    \begin{lemma}\label{lemma:PC}
    In a large society, if Assumption \ref{ass:zeal} holds, there exist thresholds, $\underline{\zeta}$, $\bar{\zeta}$, $\underline{k}_{C'}$ and $\bar{k}_{C'}$ such that if $z(f,c)\in[\underline{\zeta},\bar{\zeta}]$ and $k\in[\underline{k}_{C'},\bar{k}_{C'}]\subseteq[\underline{k}_{C},\bar{k}_{C}]$\added{, there exists a player $i\in N$ such that:}
    \begin{enumerate}[{(}i{)}]
        \item there exists a collaborative equilibrium such that $i\in C(\mathbf{g}^\ast)$ and $g^\ast_{ij}=0$ for all $j\in N$;
        \item there does not exist an independent equilibrium such that $i\in C(\mathbf{g}^\ast)$. 
    \end{enumerate}

\end{lemma}
\begin{proof}
    By Proposition \ref{prop:eq_net}, $z(f,c)\leq \bar{z}_{C}$ if a collaborative equilibrium exists. By Assumption \ref{ass:zeal}, there is an upper bound on zeal, say $\bar{\zeta}$. Moreover, 
    \begin{eqnarray*}
        GL_{i}(\mathbf{g}^\ast,\mathbf{g}^\ast+ij)&=&c(\min\{\hat{x}_i,\hat{x}_j\}+\min\{\hat{y}_i,\hat{y}_j\})+t_i\max\{[f(\hat{x}_j)-f(\hat{x}_i)],0\}\\&+&(1-t_i)\max\{[f(\hat{y}_j)-f(\hat{y}_i)],0\},
    \end{eqnarray*}
    where $(x^\ast_j,y^\ast_j)=(\hat{x}_j,\hat{y}_j)$ is decreasing in $\vert t_i-t_j\vert$ \added{as $|\hat{x}_i-\hat{x}_j|$ and $|\hat{y}_i-\hat{y}_j|$ decrease as by Lemma \ref{lemma:zeal}}(i). Thus, it is enough to consider a link to an isolated player with extreme taste \added{as more moderate types would also link to $j$ if $0$ linked to $j$}. 
\\
    Take $C(\mathbf{g}^\ast)=\{i,j\}$, with $i>j$ and suppose $g^\ast_{ij}=0$ in a collaborative equilibrium. Then, $(x^\ast_i,y^\ast_i)=(\hat{x}_i,\hat{y}_i)$ and $(x^\ast_j,y^\ast_j)=(0,\hat{y}_j-\hat{y}_i)$. Since $j\in C(\mathbf{g}^\ast)$, $c(\hat{y}_j-\hat{y}_i)\geq k$.
    \\
    If $\mathbf{g}^{\ast\ast}$ is an independent equilibrium with $\{0,i\}\in C(\mathbf{g}^{\ast\ast})$, there exists $k$ such that 
    \begin{eqnarray*}
        GL_i(\mathbf{g}^\ast,\mathbf{g^\ast}+ij)&=&c\hat{y}_i+(1-t_i)[f(\hat{y}_j-\hat{y}_i)-f(\hat{y}_i)]<k \\
        &&\leq c\hat{y}_i+(1-t_i)[f(\hat{y}_0)-f(\hat{y}_i)]=GL_i(\mathbf{g}^{\ast\ast},\mathbf{g^{\ast\ast}}+i0)
    \end{eqnarray*}
    as long as zeal is not too low, i.e., $z(f,c)\geq\underline{\zeta}$.
\\
    The independent equilibrium $\mathbf{s}^{\ast\ast}$ exists only if $k> c\hat{y}_i+(1-t_i)[f(\hat{y}_0)-f(\hat{y}_i)]=GL_i(\mathbf{g}^{\ast\ast},\mathbf{g^{\ast\ast}}+i0)$. 
    By Proposition \ref{prop:eq_net}, for a collaborative equilibrium to exist, $$GL_j(\mathbf{g}^\ast,\mathbf{g}^\ast+j0)=c(\hat{y}_0-\hat{y}_j+\hat{y}_i)<k\leq c(\hat{y}_j-\hat{y}_i)=GL_0(\mathbf{g}^\ast,\mathbf{g}^\ast+0j).$$ These conditions are consistent with $$GL_i(\mathbf{g}^\ast,\mathbf{g^\ast}+ij)=c\hat{y}_i+(1-t_i)[f(\hat{y}_j-\hat{y}_i)-f(\hat{y}_i)]<k$$ for some $k$ such that $\mathbf{s}^{\ast\ast}$ cannot be independent if the linking cost is $k$. By construction, $i$ is a potential network contributor in a collaborative equilibrium, but not in an independent equilibrium. 
    The lemma follows. 
\end{proof}
\noindent By Lemma \ref{lemma:PC}, $GL_i(\mathbf{g},\mathbf{g}+i0)=c\hat{y}_i+(1-t_i)[f(\hat{y}_0)-f(\hat{y}_i)]\geq k$ is a necessary condition for $i\in C(\mathbf{s}^W)$, where $\mathbf{s}^W$ is collaborative and $i$ has moderate taste. In the remainder of the proof, let $A$ and $B$ denote the network contributors in a collaborative equilibrium. Assume wlog $B$ free rides and $A>B$.
\begin{lemma}\label{lemma:zeal_welfare-max}
In a large society, under Assumption \ref{ass:zeal}, there exists a threshold $ z^W_{C}$, with $ z^W_{C}\leq z_{C}$, such that $z(f,c)\in[\underline{\zeta},z^W_{C}]$ is necessary for $\mathbf{s}^W$ to be collaborative.
\end{lemma}
\begin{proof}
By Proposition \ref{prop:eq_net}, $ z^W_{C}\leq z_{C}$ is a necessary condition for a collaborative equilibrium to exist. By Lemma \ref{lemma:PC}, $z(f,c)\geq\underline{\zeta}$, implying that the isolation demands of players with moderate and extreme tastes must not be too dissimilar. Hence, by \added{Lemma} \ref{lemma:zeal}.(ii), there exists $ z^W_{C}$ such that $z(f,c)\in[\underline{\zeta}, z^W_{C}]$, i.e., zeal is intermediate.
\end{proof}
%
\begin{lemma}\label{lemma:linking-cost_welfare-max}
    If Assumption \ref{ass:zeal} holds, in a large society, there exist thresholds $\underline{k}^W_{C}\geq\underline{k}_{C}$ and $\bar{k}^W_{C}\leq\bar{k}_{C}$ such that $k\in[\underline{k}^W_{C},\bar{k}^W_{C}]$ is necessary for the welfare-maximizing equilibrium to be collaborative.
\end{lemma}
\begin{proof}
By construction, $\underline{k}_{C}\leq\underline{k}^W_{C}\leq\bar{k}^W_{C}\leq\bar{k}_{C}$, since otherwise, no collaborative equilibrium exists (Proposition \ref{prop:eq_net}). Additionally, $$\added{GL_A(\mathbf{g}^\ast,\mathbf{g}^\ast+A0)=} c\hat{y}_A+(1-t_A)[f(\hat{y}_0)-f(\hat{y}_A)]\geq k,$$ \added{where $0\in C(\mathbf{g}^\ast)$ and $\mathbf{g}^\ast$ is an independent equilibrium, holds. Otherwise, there would exist an independent equilibrium where $A$ is a network contributor, a contradiction}. Hence, $\underline{k}^W_{C}$ must exist. By construction, $$LL_B(\mathbf{g}^W,\mathbf{g}^W-BA)=c(\hat{x}_B+\hat{y}_A)+t_B[f(\hat{x}_A)-f(\hat{x}_B)]\geq k$$ and $\bar{k}^W_{C}$ must exist by Lemma \ref{lemma:zeal}.(ii).
\end{proof}
\noindent The next lemma addresses the additional requirement on zeal to ensure enough moderate types benefit from $i\in C(\mathbf{g}^W)$. 
\begin{lemma}\label{lemma:xi}
    If Assumption \ref{ass:zeal} holds, in a large society with some distribution of types $\Tau$ the network in a welfare-maximizing equilibrium $\mathbf{s}^W$ is collaborative only if zeal is sufficiently low, i.e., $z(f,c)\leq\xi$.
\end{lemma}
\begin{proof}
By Lemma \ref{lemma:PC}, $A$ is a moderate type \added{who provides $(x^W_A,y^W_A)=(\hat{x}_A,\hat{y}_A)$ in a collaborative equilibrium}. Take an independent equilibrium network $\mathbf{g}^\ast$. For low enough zeal, i.e., $z(f,c)\to-1$, $C(\mathbf{g}^\ast)=\{i\}$, $x_i\to x_A$ and $y_i\to y_A$. Hence, the independent equilibrium with a single network contributor $i$ must be welfare maximizing \added{because no other network contributor can exist besides $A$.} By Lemma \ref{lemma:zeal}.(ii), if $z(f,c)\leq\xi$, $\mathbf{s}^W$ is independent. The lemma follows.
\end{proof}
\noindent \added{The condition $ z(f,c)\geq\underline{\zeta}\equiv \underline{z}^W_C$ must be satisfied together with $z(f,c)\leq\xi$ and $z(f,c)\leq\bar{\zeta}$, i.e., if $z(f,c)\leq \bar{z}_C^W\equiv\max\{\xi,\bar{\zeta}\}$.} Since there are players who can be network contributors in a collaborative equilibrium, $\mathbf{s}^W$ is collaborative under the conditions of Proposition \ref{prop:welfare-max}, which are thus also sufficient, concluding the proof.
\end{proof}

\begin{proof}[Proof of Proposition \ref{prop:welfare}] Denote by $\mathbf{s}^W(k)=(x^W(k),y^W(k),\mathbf{g}^W(k))$ the strategy profile at the corresponding welfare-maximizing equilibrium. \added{We focus on the case where the welfare-maximizing equilibrium is independent. Lemma \ref{lemma:phi} derives a necessary upper bound zeal cannot exceed. Lemma \ref{lemma:k} shows when an increase in $k$ leads to more network contributors in the welfare-maximizing equilibrium. We then use these Lemmas to show when higher linking costs increase welfare.}
\begin{lemma}\label{lemma:phi}
    If Assumption \ref{ass:zeal} holds, in a large society there exists a threshold on zeal, $\bar{z}_m$, such that there exists $\mathbf{s}^W$ with $m\in C(\mathbf{g}^W)$, where $m$ is a moderate type, only if $z(f,c)\leq\bar{z}_m$.
\end{lemma}
\begin{proof} The difference $\vert\hat{x}_i-\hat{x}_j\vert$ ($\vert\hat{y}_i-\hat{y}_j\vert$) is increasing in zeal by Lemma \ref{lemma:zeal} (ii). For some $f(\cdot)$ and $c$, $z(f,c)$ is such that, for all $j\in N\setminus\{0,1\}$, \added{if $i<j$},

$$LL_i(\mathbf{g}^\ast,\mathbf{g}^\ast-i1)=c\hat{x}_i+t_i[f(\hat{x}_1)-f(\hat{x}_i)]>c(\hat{x}_i+\hat{y}_j)+t_i[f(\hat{x}_j)-f(\hat{x}_i)]=GL_i({\bf g},{\bf g}+ij)$$ or, \added{if $i>j$}, $$LL_i(\mathbf{g}^\ast,\mathbf{g}^\ast-i0)=c\hat{y}_i+(1-t_i)[f(\hat{y}_0)-f(\hat{y}_i)]>c(\hat{x}_j+\hat{y}_i)+(1-t_i)[f(\hat{y}_j)-f(\hat{y}_i)]=GL_i({\bf g},{\bf g}+ij).$$ 
We define $\bar{z}_m$ such that this is the case whenever $z(f,c)>\bar{z}_m$. Then, there are at most two network contributors with extreme taste and $\bar{z}_m$ exists. 
\end{proof}
\begin{lemma}\label{lemma:k}
    If Assumption \ref{ass:zeal} holds, in a large society there exists a threshold $\underline{z}$, such that if $z(f,c)\in[\underline{z},\bar{ z}_m]$, there also exist thresholds $\bar{k}_2$, $\bar{k}_3$, and $\Delta$ such that if $k<\bar{k}_2<k+\Delta\leq\bar{k}_3$, $\vert C(\mathbf{g}^W(k))\vert=2$ and \added{$\vert C(\mathbf{g}^W(k+\Delta))\vert=3$}. 
\end{lemma}
\begin{proof} 
\added{Since $k<K$, $C(\mathbf{g}^W)\neq\emptyset$. There exists some threshold $\underline{z}$ such that for a moderate type $m$, if $z(f,c)<\underline{z}$, then $c\hat{x}_m\geq k>c(\hat{x}_1-\hat{x}_m)$.} So, if zeal is low enough and Assumption \ref{ass:zeal} holds, moderate players provide a lot of public goods and others link to them. \added{Moreover, the difference in total public good contributions of $m$ relative to $1$ (or $0$), is large enough such that a sufficiently high mass of free riders receive a higher utility when $C(\mathbf{g}^W)=\{m\}$. Formally, $c(\hat{x}_m+\hat{y}_m)-c\hat{x}_1\geq D$ (or $c(\hat{x}_m+\hat{y}_m)-c\hat{y}_0\geq D$), if $z(f,c)<\underline{z}$, where $D>0$. Hence, t}he welfare-maximizing equilibrium is a star if $z(f,c)<\underline{z}$. \added{We thus focus on the case where $z(f,c)\geq\underline{z}$, where there are multiple network contributors in the welfare-maximizing equilibrium.}

\noindent Fix some $i\in C(\mathbf{g}^W)$. The net gains from a link from $j$ to $i$ with $(x_i,y_i)=(\hat{x}_i,\hat{y}_i)$ are 
\begin{eqnarray*}
    GL_{j}(\mathbf{g}^\ast,\mathbf{g}^\ast+ji)-k&=&c(\min\{\hat{x}_i,\hat{x}_j\}+\min\{\hat{y}_i,\hat{y}_j\})+t_j\max\{[f(\hat{x}_i)-f(\hat{x}_j)],0\}\\
    &&+(1-t_j)\max\{[f(\hat{y}_i)-f(\hat{y}_j)],0\}-k.
\end{eqnarray*}
The expression is continuous and decreasing in $k$. For a small enough $k$ and intermediate zeal, $g^\ast_{ji}=1$ for all $j$ with $t_j\in(0,1)$. Hence, $\vert C(\mathbf{g}^W)\vert= 2$, where the other network contributor is a player with extreme taste. Since $i$ may also be a player with extreme taste, there always exists $\bar{k}_2$ such that for all $k<\bar{k}_2$, $\vert C(\mathbf{g}^W)\vert=2$.
\\
Take $C(\mathbf{g}^W)=\{A,B\}$, where $A>B$ wlog. Hence, 
\begin{eqnarray*}
    GL_A(\mathbf{g}^W,\mathbf{g}^W+AB)&=&c(\hat{x}_B+\hat{y}_A)+(1-t_A)[f(\hat{y}_B)-f(\hat{y}_A)]<k,
    \end{eqnarray*}
    \begin{eqnarray*}
    LL_1(\mathbf{g}^W,\mathbf{g}^W-1A)&=&c\hat{x}_A\geq k, \\
    GL_A(\mathbf{g}^W,\mathbf{g}^W+A1)&=&c(\hat{x}_1-\hat{x}_A)<k
\end{eqnarray*}
As $k$ increases, the network contributors are more extreme. As a result, the gains from a link to a network contributor for a moderate type $h$ are smaller, i.e., $$LL_h(\mathbf{g}^W,\mathbf{g}^W-hA)=c(\hat{x}_h+\hat{y}_A)+t_h[f(\hat{x}_A)-f(\hat{x}_h)]$$ 
is decreasing in $t_A-t_h$ while $$LL_h(\mathbf{g}^W,\mathbf{g}^W-hB)=c(\hat{x}_B+\hat{y}_h)+(1-t_h)[f(\hat{y}_B)-f(\hat{y}_h)]$$ is decreasing in $t_h-t_B$, where $A>h>B$. 
For some $k$, there exists an equilibrium $\mathbf{g}^\ast$, where $C(\mathbf{g}^\ast)=\{D,m,E\}$ and $m$ is a moderate type. If zeal is low enough, $m$ provides a lot of public good, and players with similar tastes can access a bundle close to their ideal one via a single link. By Lemma \ref{lemma:distribution}, there exists a threshold of zeal $\bar{z}_W$ such that if $z(f,c)\leq \bar{z}_W$, the isolation demand of moderate players is large enough that welfare is higher in the equilibrium with a moderate network contributor $m$, i.e., $m\in C(\mathbf{g}^W(k+\Delta)$. Defining $\bar{z}=\min\{\bar{z}_m,\bar{z}_W\}$, the lemma follows.
\end{proof}

\noindent Suppose $C(\mathbf{g}^W(k))=\{A,B\}$, with $A>B$. Take $k$ and $\Delta$ such that $C(\mathbf{g}^W(k+\Delta))=\{D,m,E\}$, with $D>m>E$. With some abuse of notation, denote by $t_A$ ($t_B$) the lowest (highest) type who links to $A$ ($B$), with $t_A\leq t_B$. Denote by $\bar{t}_m$ ($\underline{t}_m$) the highest (lowest) type who links to $m$. Take player $i$, for whom $g^W_{im}(k+\Delta)=1$. We distinguish two cases and assume $i>m$ wlog.
\\
\textbf{Case 1}: $g^W_{iA}(k)=g^W_{iB}(k)=1$. By Proposition \ref{prop:charact}, $A>i>B$. Then, $U_i(\mathbf{s}^W(k))<U_i(\mathbf{s}^W(k+\Delta))$ yields $$\Delta<k-c(\hat{x}_i-\hat{x}_m)-t_i[f(\hat{x}_A+\hat{x}_B)-f(\hat{x}_i)]-(1-t_i)[f(\hat{y}_A+\hat{y}_B)-f(\hat{y}_m)].$$ \added{Note that player $i$ is active in $x$ in $\mathbf{s}^W(k+\Delta)$, but is inactive in $\mathbf{s}^W(k)$.}
\\
\textbf{Case 2}: $g^W_{iA}(k)=1$ and $g^W_{iB}(k)=0$. Then, $U_i(\mathbf{s}^W(k))<U_i(\mathbf{s}^W(k+\Delta))$ yields $$\Delta<c(\hat{y}_i-\hat{y}_A-(\hat{x}_i-\hat{x}_m))-t_i[f(\hat{x}_A)-f(\hat{x}_i)]+(1-t_i)[f(\hat{y}_m)-f(\hat{y}_i)].$$ \added{Note that player $i$ is active in $y$ in $\mathbf{s}^W(k)$ and active in $x$ in $\mathbf{s}^W(k+\Delta)$.}
\\
If $t_i$ is close to $m$, the RHS of the two inequalities is strictly positive. We can therefore find $\Delta$ that satisfies them. If additionally, zeal is sufficiently low, i.e., $z(f,c)\leq\zeta$, by Lemma \ref{lemma:distribution}, then the conditions hold for more types and $\sum_{i\in N}u(\mathbf{s}^W(k))<\sum_{i\in N}u(\mathbf{s}^W(k+\Delta))$. Redefine then the threshold $\bar{z}$ from the previous lemma as follows: $\bar{z}=\min\{\bar{z},\zeta\}$, i.e., the lower of the two threshold on zeal is binding and implies the other condition. This concludes the proof of Proposition \ref{prop:welfare}.
\end{proof}

\begin{proof}[Proof of Proposition \ref{prop:polarization}] Denote by $\mathbf{s}^W(k)=(x^W(k),y^W(k),\mathbf{g}^W(k))$ the strategy profile at the corresponding welfare-maximizing equilibrium. \added{We first derive in Lemma \ref{lemma:thresholds} the linking costs such that there are at most two network contributors.}
\begin{lemma}\label{lemma:thresholds}
    If Assumption \ref{ass:zeal} holds, in a large society and $z(f,c)\neq0$, there exist thresholds $\underline{k}$ and $\bar{k}$, with $0<\underline{k}\leq\bar{k}<K$ such that $\vert C(\mathbf{g}^W)\vert\leq2$ if $k\in(0,\underline{k}]$ or $k\in[\bar{k},K]$, and $\mathbf{g}^W$ is independent.
\end{lemma}
\begin{proof} By Proposition \ref{prop:eq_net}, no collaborative equilibria exist if $k$ is sufficiently high or low. If $k\to0$, for all $i\in N\setminus\{0,1\}$, either $g^W_{i1}=1$, or $g^W_{i0}=1$, or both. Indeed, $\hat{x}_1>\hat{x}_i$ and $\hat{y}_0>\hat{y}_j$, so $i$ links to $1$ and $0$ for low enough $k$. Hence, $C(\mathbf{g}^W)=\{A,B\}$ for $k\in(0,\underline{k}]$. If $k>K$, the empty network is the unique equilibrium. If $z(f,c)>0$, by Lemma \ref{ass:zeal}.(ii), there exists $\bar{k}$ such that for $k\in[\bar{k},K]$, $C(\mathbf{g}^W)=\{A,B\}$ as players with more extreme taste provide more total public good (Lemma \ref{lemma:zeal}.(i)).
\end{proof}

\noindent We study the effects on polarization of an increase of $\Delta>0$ in the linking cost $k$ so that $k,k+\Delta\in (0,\underline{k}]$ or $k,k+\Delta\in [\bar{k},K]$ if additionally zeal is positive, so that $C(\mathbf{g}^W(k))=\{A,B\}$ and $C(\mathbf{g}^W(k+\Delta))=\{A',B'\}$, where $A>B$ is wlog. 
\\
By Proposition \ref{prop:charact} we can then partition the type space into: 
\begin{enumerate}
    \item $[0,\underline{t}_A]$ and $[0,\underline{t}_{A'}]$, i.e., types who only link to 
\added{$B$ in $\mathbf{g}^W(k)$ and to $B'$ in $\mathbf{g}^W(k+\Delta)$};
\item \added{$(\underline{t}_A,\bar{t}_B)$ and $(\underline{t}_{A'},\bar{t}_{B'})$, i.e., types who link to $A$ and $B$ in $\mathbf{g}^W(k)$ and to $A'$ and $B'$ in $\mathbf{g}^W(k+\Delta)$);}
\item \added{$[\bar{t}_B,1]$ and $[\bar{t}_{B'},1]$, i.e., types who only link to 
$A$ in $\mathbf{g}^W(k)$ and to $A'$ $\mathbf{g}^W(k+\Delta)$);}
\item \added{$(\bar{t}_B,\underline{t}_A)$ and $(\bar{t}_{B'},\underline{t}_{A'})$, i.e., isolated players both in $\mathbf{g}^W(k)$ and in $\mathbf{g}^W(k+\Delta)$.}
\end{enumerate}
\added{By Proposition \ref{prop:charact} if some players link to both $A$ and $B$, there are no isolated players, so cases 2 and 4 are mutually exclusive.} We distinguish the two cases.
\\
\textbf{Case 1:} Low $k$---no isolated players. By Proposition \ref{prop:charact}, $\underline{t}_A<\underline{t}_{A'}<\bar{t}_{B'}<\bar{t}_B$. As $GL_i(\mathbf{g},\mathbf{g}+ij)$ is strictly concave in $(\hat{x}_i,\infty)$ and $(\hat{y}_i,\infty)$, this also holds if $A\geq A'>B'\geq B$. The change in polarization among agents with different linking strategies is:
{\small
\begin{eqnarray}
&&\int^{1}_{\underline{t}_{A'}}\int^{\underline{t}_{A'}}_0(\vert \bar{x}^W_i(k+\Delta) -\bar{x}_j^W(k+\Delta) \vert + \vert \bar{y}^W_i(k+\Delta)-\bar{y}_j^W(k+\Delta)\vert) d\tau(i)d\tau(j) \notag \\
&& +\int^{\bar{t}_{B'}}_0\int^{1}_{\bar{t}_{B'}}(\vert \bar{x}^W_i(k+\Delta)-\bar{x}_j^W(k+\Delta) \vert + \vert \bar{y}^W_i(k+\Delta)-\bar{y}_j^W(k+\Delta) \vert) d\tau(i)d\tau(j)\notag \\
&& - \int^{1}_{\underline{t}_A}\int^{\underline{t}_A}_0(\vert \bar{x}^W_i(k)-\bar{x}_j^W(k) \vert + \vert \bar{y}^W_i(k)-\bar{y}_j^W(k)\vert) d\tau(i)d\tau(j)\notag \\
&& - \int^{\bar{t}_B}_0\int^{1}_{\bar{t}_B}(\vert \bar{x}^W_i(k)-\bar{x}_j^W(k) \vert +
\vert \bar{y}^W_i(k)-\bar{y}_j^W(k) \vert) d\tau(i)d\tau(j), \label{eq:polarization2_final_case1}
\end{eqnarray}}
where $\bar{x}_i^W(k)$ is either $\hat{x}_i$, $\hat{x}_A$, or $\hat{x}A+\hat{x}B$, depending on whether $i>A$ or $i<A$ and whether $i$ sponsors one or two links in $\mathbf{s}^W(k)$. For sufficiently low $k$, \eqref{eq:polarization2_final_case1} is larger, since fewer players sponsor links as $k$ increases. By Lemma~\ref{lemma:zeal}, $\hat{x}_{i}-\hat{x}_j>0$ ($\hat{y}_{i}-\hat{y}_{j}>0$) is increasing in $|t_i-t_j|$ for $i>j$ ($i<j$). Hence, polarization increases both across players with different linking strategies and among those with the same strategy, implying a strict increase in polarization. This proves Proposition \ref{prop:polarization}.$(i)$.

\noindent \textbf{Case 2}: High $k$, i.e., isolated players. If $z(f,c)>0$, isolated players are of moderate taste. Clearly, $A'\geq A>B\geq B'$, since either $g^W_{1A'}(k+\Delta)=1$ ($g^W_{0B'}(k+\Delta)=1$) or $A'=1$ ($B'=0$). Then, $\underline{t}_{A'}>\underline{t}_{A}>\bar{t}_B>\bar{t}_{B'}$.
\\
The change in polarization among players who have the same linking strategy is
{\small
\begin{eqnarray}
&& \int^{\underline{t}_{A'}}_0\int^{\underline{t}_{A'}}_0(\vert \bar{x}^W_i(k+\Delta)-\bar{x}^W_j(k+\Delta)\vert+\vert \bar{y}_i^W(k+\Delta) -\bar{y}^W_j(k+\Delta)\vert)d\tau(i)d\tau(j)\notag \\
&& +\int^{1}_{\bar{t}_{B'}}\int^{1}_{\bar{t}_{B'}}(\vert\bar{x}_i^W(k+\Delta )-\bar{x}^W_j(k+\Delta)\vert+\vert\bar{y}^W_i(k+\Delta)-\bar{y}^W_j(k+\Delta)\vert)d\tau(i)d\tau(j) \notag \\
&& -\int^{\underline{t}_A}_0\int^{\underline{t}_A}_0(\vert \bar{x}^W_i(k)-\bar{x}^W_j(k)\vert+\vert\bar{y}_i^W(k)-\bar{y}^W_j(k)\vert) d\tau(i)d\tau(j)\notag \\
&& -\int^{1}_{\bar{t}_B}\int^{1}_{\bar{t}_B}(\vert\bar{x}_i^W(k)-\bar{x}^W_j(k)\vert+\vert\bar{y}^W_i(k)-\bar{y}^W_j(k) \vert )d\tau(i)d\tau(j),
\label{eq:polarization2_final}
\end{eqnarray}}
where $\bar{x}_i^W(k)$ is either $\hat{x}_i$ or $\hat{x}_A$, depending on whether $i>A$ or $i<A$. If $k$ is high enough, \eqref{eq:polarization2_final} is lower, as more players are isolated. Polarization decreases when zeal is not too high (Lemma \ref{lemma:distribution}). \added{Formally, $z(f,c)<\xi$.} As polarization in all other partitions strictly decreases, polarization strictly decreases. This concludes the proof of Proposition \ref{prop:polarization}.
\end{proof}

\begin{proof}[Proof of Proposition \ref{prop:subsidy}]
The social planner targets the subsidy to a player receiving links in the welfare-maximizing equilibrium after the subsidies, $\mathbf{g}^{W(\mathbf{v})}$, as this generates consumption spillovers, hence larger welfare. 
\\
Take an equilibrium $\mathbf{s}^\ast$, with $i\in C(\mathbf{g}^\ast)$, and $j$. Let $j\not\in C(\mathbf{g}^\ast)$, $g^\ast_{ji}=1$ and $g^\ast_{hj}=0$ for all $h\in N$. Hence, $j$ is a free rider in $\mathbf{g}^{\ast(0)}$ (the case of no subsidy). As $v_j>0$ only if $j\in C(\mathbf{g}^{\ast(\mathbf{v})})$, $c(x_j^{\ast(\mathbf{v})}+y_j^{\ast(\mathbf{v})})\geq k$, which requires $V\geq v_j(x_j^{\ast(\mathbf{v})}+y_j^{\ast(\mathbf{v})})$. Hence, there exists $\underline{V}$, such that for $V\leq \underline{V}$, $v_i>0$ if and only if $i\in C(\mathbf{g}^\ast)$.
\\
For $V>\underline{V}$, there is a player $m$ such that $v_m>0$, $m\in C(\mathbf{g}^{\ast(\mathbf{v})})$ but $m\notin C(\mathbf{g}^\ast)$. If $GL_i(\mathbf{g}^{\ast(\mathbf{v})},\mathbf{g}^{\ast(\mathbf{v})}+im)
<k$ for some $i\in N$, $\vert C(\mathbf{g}^{\ast(\mathbf{v})})\vert\geq 2$. Otherwise, $C(\mathbf{g}^{\ast(\mathbf{v})})=\{m\}$. Since 
\begin{eqnarray*}
GL_i(\mathbf{g}^{\ast(\mathbf{v})},\mathbf{g}^{\ast(\mathbf{v})}+im)&=&c(\min\{\hat{x}_m^{\ast(\mathbf{v})},\hat{x}_i\}+\min\{\hat{y}_m^{\ast(\mathbf{v})},\hat{y}_i\})+t_i\max\{[f(\hat{x}_m^{\ast(\mathbf{v})})-f(\hat{x}_i)],0\}\\&+&(1-t_i)\max\{[f(\hat{y}_m^{\ast(\mathbf{v})})-f(\hat{x}_i)],0\},
\end{eqnarray*}
it is increasing in $v_m$; hence, there exists $\overline{V}$ such that $\vert C(\mathbf{g}^{\ast(\mathbf{v})})\vert\geq 2$ for $V\in[\underline{V},\overline{V})$ and $C(\mathbf{g}^{\ast(\mathbf{v})})=\{m\}$ for $V\geq \overline{V}$. This concludes the proof of Proposition \ref{prop:subsidy}.
\end{proof}

\begin{proof}[Proof of Corollary~\ref{cor:linksubsidy}]
\added{
Because link subsidies do not affect players’ public-good provision, their only effect is to reduce the effective cost of forming links. If equilibrium linking strategies are unchanged, the result follows immediately. 
Consider the subsidized equilibrium $\mathbf{g}^{W(\mathbf{l})}$. Links in this equilibrium fall into two categories.
}
\\
\added{First, links that formed even without subsidies. For these links, $GL_j(\mathbf{g}^{W(\mathbf{0})},\mathbf{g}^{W(\mathbf{0})}+ji)\ge k$, so any subsidy $l_{ji}$ paid on such links is a pure transfer and 
does not increase welfare. Second, links induced by subsidies. For any such link $(j,i)$,
\[
GL_j(\mathbf{g}^{W(\mathbf{l})},\mathbf{g}^{W(\mathbf{l})}+ji)\ge k-l_{ji}
\quad\text{but}\quad
GL_j(\mathbf{g}^{W(\mathbf{0})},\mathbf{g}^{W(\mathbf{0})}+ji)<k.
\]
If player $j$ is exactly indifferent absent the subsidy, then
$GL_j(\mathbf{g}^{W(\mathbf{l})},\mathbf{g}^{W(\mathbf{l})}+ji)=k-l_{ji}$,
so the net private gain from the link is zero and the welfare gain equals the subsidy $l_{ji}$.
If the link is strictly profitable due to the subsidy, the additional gain arises solely from the reduction in the linear linking cost and is therefore at most $l_{ji}$. 
Summing across all links, the total welfare gain is at most $L(\mathbf{l})$, which concludes the proof.}
\end{proof}

\end{appendices}

\end{document}